\documentclass[a4paper,12pt]{report}
\usepackage{amsfonts,amsmath,amssymb}
\usepackage[latin1,cp1251]{inputenc}
\usepackage[russian]{babel}
\usepackage[dvips]{graphicx}
\begin{document}
\def\rot{\mathop{\rm rot}\nolimits}
\def\div{\mathop{\rm div}\nolimits}
\def\grad{\mathop{\rm grad}\nolimits}
\def\gv#1{\mathop{\vspace{2pt}\mathfrak  #1}\nolimits}
\def\rv#1{\mathop{\vspace{2pt}\rm \bf #1}\nolimits}
\newcommand{\Arctg}{\mathop{\rm Arctg}\nolimits}
\newcommand{\Arcctg}{\mathop{\rm Arcctg}\nolimits}
\newcommand{\re}{\mathop{\rm Re}\nolimits}
\newcommand{\im}{\mathop{\rm Im}\nolimits}
\newenvironment{fr}[1]{\begin{tabular}{|p{#1}|}\hline}
     {\\\hline\end{tabular}}
\newcommand{\fg}[2]{\par\vspace{-\parskip}\hspace*{-22pt}
     \hangindent=#1\hangafter=#2}
\newcommand{\sign}{\mathop{\rm sign}\nolimits}

\thispagestyle{empty}
%\baselineskip=\normalbaselineskip
%\baselineskip=1.005\normalbaselineskip
\begin{center}\bf{Учреждение Российской академии наук\\ИНСТИТУТ ЯДЕРНЫХ ИССЛЕДОВАНИЙ РАН}\end{center}
\vspace*{35mm}
\begin{center}\large\textbf{В.А. Бережной$^{\bf 1,2}$,\hspace{0.25cm} В.Н. Курдюмов$^{\bf 3}$}
\end{center}
\vspace*{5mm}
\begin{center}\Large{\bf Излучение точечного заряда, движущегося \\[2mm]
равномерно вдоль оси круглого\\[2mm] диафрагмированного волновода}
\end{center} 

\vspace{10mm} ${}^{\bf 1}$ {Учреждение Российской академии наук Институт ядерных исследований РАН}

\vspace*{1mm} ${}^{\bf 2}$ {Национальный исследовательский университет <<МФТИ>>}

\vspace*{1mm} ${}^{\bf 3}$ {Московский радиотехнический институт РАН}
%\vspace*{20mm}\\
%\vspace*{2mm}
%В печать, в свет\\\vspace*{3mm}
%Директор ИЯИ РАН\\\vspace*{3mm}
%академик \rule{35mm}{0.4pt} В.А. Матвеев\\\vspace*{2mm}
%\rule{15mm}{0.4pt}  июня 2010 г.
%\vspace*{20mm}
\vspace*{75mm}
\begin{center}\normalsize{МОСКВА\hspace{0.75cm}2010\hspace{0.75cm}MOSCOW}
\end{center}
\newpage
%\end{document}
\thispagestyle{empty}
УДК 537.876; 621.384.6 \vspace*{1cm}
\begin{center}
           {\bf{Излучение точечного заряда,  \\[0.05mm]движущегося
равномерно вдоль оси \\ [0.05mm]круглого диафрагмированного волновода}}\vspace*{0.5cm}
      \begin{minipage}[c]{0.75\textwidth}
         \parindent0.75cm
      {\small {Проанализировано пространственно-временное распределение поля, возбужд\"енного зарядом,
      движущимся по оси круглого диафрагмированного волновода. Численно определены потери энергии
      при прол\"ете зарядом периода структуры; показано, что при увеличении релятивистского фактора все параметры каждой собственной волны стремятся к конечным значениям, определяющим предельный спектр излучения. С ростом энергии заряда происходит насыщение амплитуд нижней части спектра собственных волн на уровне предельного спектра. Для получения и интерпретации результатов используется принцип
      трансляционной симметрии поля. }}      
      \end{minipage}
\end{center}\vspace*{1cm}
\begin{center}
           {\bf{Radiation of a point charge \\[0.05mm]uniformly moving 
            along the axis \\[0.05mm]of the circular disc-loaded waveguide}}\vspace*{0.5cm}
      \begin{minipage}[c]{0.75\textwidth}
         \parindent0.75cm
      {\small { The space-charge distribution of the electromagnetic field excited by point charge moving with a constant velocity on the axis of the circular disc-loaded waveguide was analyzed. The energy losses per structure period numerically were calculated. With increasing charge energy main features of the waveguide modes reach some finite values and, so way, define the limit radiation spectrum. The saturation of the amplitudes of modes (which belongs to the lower part of the limit spectrum) was observed. The analytical techniques and results of calculations were discussed with the use of translation symmetry and the problems that should be noticed were pointed out. }}
      \end{minipage}
\end{center}
\vspace*{3cm} \hspace*{6.1cm}$\copyright$ Учреждение Российской академии наук

 \hspace*{6.1cm} Институт  ядерных исследований РАН, 2010

\newpage\vspace*{15mm}
   \begin{center}
   \large\textbf{Введение}
   \end{center}

   В настоящем препринте излагаются результаты расч\"ета
   электромагнитного поля, возбуждаемого точечным зарядом при его
   равномерном движении вдоль оси неограниченной
   аксиально-симметричной замедляющей периодической структуры, в
   качестве которой выбран круглый диафрагмированный волновод (КДВ).
   Отрезки такого волновода находят широкое применение в качестве
   ускоряющих секций современных линейных ускорителей электронов.

   Аналитическая теория собственных волн КДВ к
   настоящему времени достаточно хорошо разработана [1], во всяком
   случае достигнуто понимание структуры их низкочастотной части спектра 
   и основных физических свойств. Все строгие количественные способы
   расч\"ета характеристик этих волн используют метод частичных
   областей (МЧО), который позволяет свести задачу нахождения
   распределения поля в КДВ к бесконечной системе линейных
   алгебраических уравнений для амплитуд парциальных волн простых
   регулярных областей. Решение таких систем подразумевает их
   редукцию и использование для расч\"етов компьютеров.
   Достигаемая при этом точность вычисления спектра нижней части
   спектра во всяком случае превышает потребности практики при
   современном уровне технологии изготовления секций волновода; в
   этой работе используется редко встречающийся в теории КДВ
   вариант МЧО [2,3].

   Теория возбуждения КДВ движущимися заряженными источниками
   разработана несравненно хуже. Хотя и имеется ряд работ [4,5], в
   которых рассматривается такая задача, однако в них отсутствует
   анализ пространственно-временного распределения возбуждаемого
   поля и нет количественных результатов вычисления потерь
   энергии источниками для близких к используемым на практике
   размерам КДВ. Более того, в этих работах встречаются ряд
   противоречащих друг другу и просто ошибочных заключений. К
   таким, например, можно отнести утверждение, что впереди
   заряда возбуждаются волны с положительной групповой скоростью,
   а сзади --- с отрицательной.

   Одной из основных причин допущенных в ряде работ
   неточностей является недостаточно обоснованный и формальный
   перенос разработанной Л.А. Вайн\-штейн\-ом [6] теории возбуждения
   однородных волноводов на случай периодических структур. На
   возможность такого обобщения указывалось и в работах автора
   теории [6,7]. Во избежание ряда трудностей, присущих этому
   методу, ниже рассматривается решение задачи, полученное не
   столь общим методом, имеющим, однако, то преимущество, что его
   применимость в данном конкретном случае не требует
   дополнительных обоснований.

   Поскольку далее ищется стационарное решение, то в его основу
   должен быть положен принцип трансляционной симметрии, который
   требует, чтобы при перемещении заряда на целое число
   пространственных периодов структуры вся картина распределения
   электромагнитного поля сместилась на столько же периодов.
   Вопрос о корректности постановки задачи для неограниченной
   структуры и существовании у не\"е единственного решения
   требует, по-видимому, дополнительного обоснования. Для нас же
   будет достаточно того, чтобы решение удовлетворяло некоторым
   общим физическим требованиям.

   Необходимо, чтобы выполнялся закон сохранения энергии:
   изменение энергии электромагнитного поля в структуре должно
   быть равно работе этого поля над зарядом в области, которую заряд 
   уже прош\"ел. Отсюда с
   неизбежностью следует, что возбуждаемое зарядом поле должно
   простираться вдоль КДВ до бесконечности без спадания амплитуды,
   то есть иметь волновой характер. В противном случае имел бы
   место трансляционный перенос поля вместе с зарядом, не
   происходило бы изменения его энергии, а значит и торможения
   заряда, что противоречит хорошо известным фактам.

   Вообще говоря, поле может иметь волновой характер как впереди
    (по ходу движения), так и позади заряда. В ряде работ именно
   эта возможность молчаливо подразумевается авторами. Нам
   представляется очевидным, что решение задачи об излучении
   заряда в неограниченном КДВ может иметь практическое приложение
   к реальным структурам только в том случае, когда поле на
   большом расстоянии впереди заряда не имеет волнового характера
   и убывает с расстоянием от заряда как обычное кулоновское поле
   в регулярной структуре, ограниченной в поперечном
   направлении. Поэтому естественно  ожидать, что только в той области КДВ,
   которую заряд уже пролетел, поле имеет волновой характер и представляет
    собой суперпозицию собственных волн КДВ с постоянными амплитудами.

   Полное изменение энергии поля в объ\"еме КДВ при перемещении
   заряда, а, следовательно, и теряемая им кинетическая энергия,
   при условии постоянства амплитуд возбужд\"енных собственных волн
   обусловлено неизбежным изменением объ\"ема структуры, заполняемого полем
   по мере следования заряда, и может дополняться ещ\"е
   потоком энергии через поперечное сечение на бесконечность.
   Представляется естественным, что этот поток должен быть
   направлен от заряда и, следовательно, в волновом поле преобладающий
    вклад в этот поток должны давать собственные волны с отрицательной
    групповой скоростью, у    которых направление вектора Пойнтинга
    противоположно направлению набега фазы волны на периоде (последнее в силу
   трансляционной симметрии совпадает с направлением движения
   заряда). В противном случаен в КДВ имел бы место стационарный
   поток энергии из области, где отсутствуют источники излучения.\\

   \begin{center}
   \large\textbf{1. Поле точечного заряда, движущегося по оси круглого
   диафрагмированного волновода}
   \end{center}

      Пусть точечный заряд \(Q\) движется с постоянной скоростью
      \(v=\beta c\) вдоль оси КДВ. Обозначение всех размеров
      волновода и расположение используемой далее цилиндрической
      системы координат \((\rho, \varphi,z)\) ясны из рис. 1. Из-за симметрии структуры
      и траектории движения заряда следует, что в возбуждаемом электромагнитном поле
      отличные от нуля составляющие  \(E_{\rho}, E_z, H_{\varphi}\) не зависят от угла \(\varphi\).
      Поле заряда должно обладать  трансляционной симметрией, то есть все его составляющие
       удовлетворяют соотношениям вида
      \[ E_z(\rho,z+2Dn,t+2Dn/v)=E_z(\rho,z,t)\,,\eqno(1.1)\]
      где \(n\) -- число пространственных периодов КДВ (\(n<0\) -- слева, \(n>0\) -- справа от
      плоскости \(z=0\)). Поэтому  достаточно найти  поле   в какой-нибудь одной ячейке, представляющей
      собой период КДВ:  поле в любой другой ячейке, отстоящей на \(n\) периодов, будет
      таким же, но через интервал времени  \( \Delta t= 2Dn/ v\).
       Выберем в качестве такой ячейки область \(-d<z\leqslant 2D-d\),
      в которой можно выделить две  частичные области,
      представляющие собой отрезки круглого волновода соответствующего
      радиуса: область \(A\, (d<z\leqslant 2D-d,\,0<\rho<a)\) и область \(B\,(|z|<d,\,0<\rho<b)\).

   \hspace{2.0cm}\includegraphics[bb= 0mm 0mm 400mm 300mm,scale=0.23,draft=false]{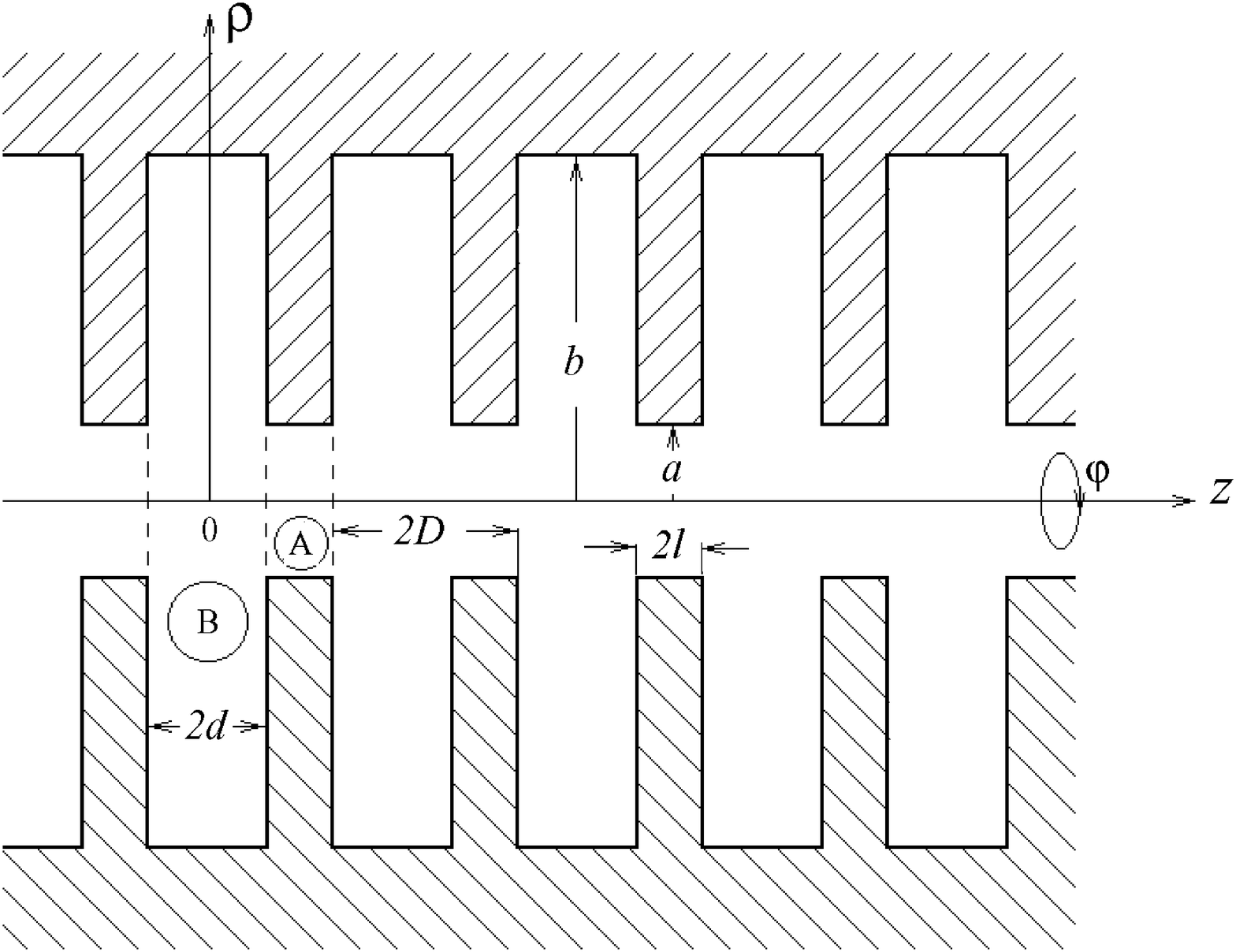}

    \vspace{0.1cm}
    \hspace{15mm}Рис.1. Сечение круглого диафрагмированного волновода
    \vspace{0.2cm}

            В поле заряда, движущегося равномерно вдоль оси однородного круглого волновода
       радиуса \(R\) и пролетающего плоскость \(z=0\) в момент \(t=0\), отличны от нуля составляющие
       \[\left.\begin{array}{lcl}E_z^{R}(\rv r,t) &
         = &\displaystyle{\frac{2Q}{R^2}\sign{(z\!-\!vt)}\sum_{n=1}^\infty \frac{J_0(\nu_n\rho/R)}
         {J_1^2(\nu_n)}\,e^{-|z-vt|\nu_n \gamma/R}}\,,\\[0.7cm]
         E_{\rho}^{R}(\rv r,t) &
         = &\displaystyle{\frac{2Q}{R^2}\gamma\sum_{n=1}^\infty \frac{J_1(\nu_n\rho/R)}
         {J_1^2(\nu_n)}\,e^{-|z-vt|\nu_n \gamma/R}}\,,\\[0.7cm]
         H_\varphi^{R}(\rv r,t)&=& \beta E_\rho^R(\rv r,t)\,,
         \end{array}\right \}\eqno(1.2)\]
         где релятивистский фактор \(\gamma=1/{\sqrt{1-\beta^2}}\), \(J_n(x)\) -- функции Бесселя порядка \(n\),
         \(\nu_n\) -- корни  уравнения \(J_0(x)=0\)(см., например, [8]).

         Раскладывая все составляющие поля в интегралы Фурье вида
         \[\rv E(\rv r,t)=\int\limits_{-\infty}^\infty \rv E(\rv r,\omega)\,
            e^{-i\omega t}\,\mathrm{d} \omega \,,\eqno(1.3)\]
           получаем следующие выражения для Фурье-компонент поля (1.2):\\
           \[\left.\begin{array}{rclcl}E_z^{R}(\rv r,\omega) &
         = &-i\displaystyle{\frac{Q\Gamma}{\pi v\gamma}\Big[K_0(|\Gamma|\rho)-
         I_0(\Gamma\rho)\frac{K_0(|\Gamma |R)}{I_0(\Gamma R)}\Big ]\,
         e^{i\frac \omega v z}}&=&E_z^{R}(\rho,\omega)e^{i\frac \omega v z}\,,\\[.75 cm]
         E_\rho^{R}(\rv r,\omega) & = &\displaystyle{\phantom{-i}\frac{Q\Gamma}
         {\pi v}\Big[K_1(|\Gamma|\rho)+I_1(\Gamma\rho)\frac{K_0(|\Gamma |R)}
         {I_0(\Gamma R)}\Big ]}\,e^{i\frac \omega v z}&=
         &E_\rho^{R}(\rho,\omega)e^{i\frac \omega v z}\,,\\[.75 cm]
         H_\varphi^{R}(\rv r,\omega)&=& \beta E_\rho(\rv r,\omega)\,,
         \end{array}\right \}\eqno(1.4)\]
         где \(\Gamma=k/{\beta \gamma},\; k=\omega/c,\;
        K_n(x) \) и \(I_n(x)\) ---  соответственно функции Макдональда и модифицированные
        функции Бесселя 1-го рода порядка \(n\).

      Совокупность полей (1.2) в двух частичных областях
      рассматриваемой ячейки, очевидно, не может служить решением
      задачи, поскольку при этом на границе областей в плоскости \(z=d\) не выполняется
      граничное условие \(E_\rho=0\) в интервале \(a<\rho<b\) и не
      обеспечена непрерывность тангенциальных составляющих полей
      при \(0<\rho<a\). Для выполнения перечисленных условий в каждой
      из выделенных областей ячейки к Фурье-компонентам (1.4) необходимо
      добавить Фурье-компоненты поля, удовлетворяющего уравнениям Максвелла,
      свободным от источников. С учетом поляризации поля движущегося заряда
      в качестве такой добавки достаточно взять суперпозицию азимутально симметричных
      \(E-\)волн круглого волновода соответствующего радиуса \(R\):
      \[ \left.\begin{array}{lll} E_{\pm s,\,\rho}^R (\rv r,\omega)&=& \pm\,\displaystyle{\frac
         {h^R_s\,J_1(\nu_s\rho/R)}{R\,J_1(\nu_s)}}\,e^{\pm ih_s^R z},\\
         [.75 cm]H_{\pm s,\,\varphi}^R(\rv r,\omega) &=& \phantom{\pm}\displaystyle{\frac
         {k\,J_1(\nu_s\rho/R)}{R\,J_1(\nu_s)}}\,e^{\pm ih_s^R z},\\
         [.75 cm]E_{\pm s,\,z}^R(\rv r,\omega) &=& i\;\displaystyle{\frac
         {\nu_s\,J_0(\nu_s\rho/R)}{R^2\,J_1(\nu_s)}}\,e^{\pm ih_s^R z}\,,\\
         [.75 cm]\end {array}\right \}\qquad s=1,2,\ldots\,,
         \eqno(1.5)\]
   где продольные волновые числа \(h_s^R= \sqrt{k^2- (\nu_s/R)^2}\).
   Выбранная нормировка функций (1.5)  делает наиболее компактной запись последующих уравнений.

     Итак, Фурье-компоненты поля в двух регулярных областях ячейки ищутся в виде:

     в области\(A\)  ---
    \[ \left. \begin{array}{lll} E_\rho(\rv r,\omega) & = & \sum\limits_{s=1}^\infty
    \big[ A_s E_{s,\,\rho}^a(\rv r,\omega)+A_{-s}
     E_{-s,\,\rho}^a(\rv r,\omega)\big] + E_\rho^a(\rv r,\omega)\,,\\[0.6cm]
     E_z(\rv r,\omega) & = & \sum\limits_{s=1}^\infty \big[A_s E_{s,\,z}^a(\rv r,\omega)+A_{-s}
     E_{-s,\,z}^a(\rv r,\omega) \big]+ E_z^a(\rv r,\omega)\,,\\[0.6cm] H_\varphi(\rv
     r,\omega) &=& \sum\limits_{s=1}^\infty\big[ A_s H_{s,\,\varphi}^a(\rv r,\omega)+A_{-s}
     H_{-s,\,\varphi}^a (\rv r,\omega)\big]+ \beta
     E_\rho^a(\rv r,\omega)\,;\\[0.3cm]\end{array}\right\}
      \eqno(1.6)\]

     в области\(B\)  ---
          \[\left. \begin{array}{lll} E_\rho(\rv r,\omega) & = & \sum\limits_{s=1}^\infty
       \big[ B_s E_{s,\,\rho}^b(\rv r,\omega)
       +B_{-s}  E_{-s,\,\rho}^b(\rv r,\omega)\big] + E_\rho^b(\rv r,\omega)\,,\\[0.6cm] E_z(\rv
     r,\omega)& = & \sum\limits_{s=1}^\infty \big[B_s E_{s,\,z}^b(\rv r,\omega)+B_{-s}
     E_{-s,\,z}^b(\rv r,\omega)\big] + E_z^b(\rv r,\omega)\,,\\[0.6cm] H_\varphi(\rv
     r,\omega)&=& \sum\limits_{s=1}^\infty \big[B_s H_{s,\,\varphi}^b(\rv r,\omega)+B_{-s}
     H_{-s,\,\varphi}^b(\rv r,\omega)\big] + \beta
     E_r^b(\rv r,\omega)\,,\\[0.3cm]\end{array}\right\}\eqno(1.7)\]
где \(A_s, A_{-s}, B_s, B_{-s}\) -- искомые коэффициенты, зависящие при заданных размерах
КДВ и параметрах заряда только от частоты \(\omega\), а остальные функции в правых
частях этих соотношений определены выше.

        Условие непрерывности поперечных составляющих полей на
       уступе круглого волновода в плоскости \(z=d\) (при \(\rho<a\)) и требование
       обращения в нуль тангенциальной составляющей электрического
       поля на торцевой стенке (при \(a<\rho<b\)) приводит к следующим
       функциональным уравнениям относительно переменной \(\rho\):

      для электрического поля
         \[\frac 1 b\sum_{n=1}^\infty\frac{J_1(\nu_n \rho/b)}{J_1(\nu_n
         )}h_n^b\left [B_ne^{ih_n ^b d}-B_{-n}e^{-ih_n^b d}\right
         ]+E^b_\rho(\rho,\omega)e^{i\frac \omega v d}=\\[.4cm]\]
     \vspace{-1.2cm}\[\eqno(1.8)\]\vspace{-0.5cm}
     \[\hspace{1cm} =\left\{\begin{array}{ll}0&\mbox{при~\(a<\rho<b,\)}
         \\\displaystyle{\frac 1 a\sum_{n=1}^\infty\frac{J_1(\nu_n \rho/a)}{J_1
         (\nu_n)}h_n^a}\left [A_n e^{ih_n^a d}-A_{-n}e^{-ih_n^a d}
         \right ]+\!E^a_\rho(\rho,\omega)e^{i\frac \omega v d}
         &\mbox{при~\(0<\rho<a;\)}\end{array}\right.\]
\vspace{1.5mm}

      для магнитного поля

        \[\frac k b\sum_{n=1}^\infty \frac{J_1(\nu_n \rho/b)}{J_1(\nu_n)}
        \left [B_ne^{ih_n^b d}+B_{-n}e^{-ih_n^b d}\right ]+\beta
        E^b_\rho(\rho,\omega)e^{i\frac \omega v d}=\\[.4cm]\]
     \vspace{-0.9cm}\[\eqno(1.9)\]\vspace{-0.5cm}
     \[=\displaystyle{\frac k a\sum_{n=1}^\infty\frac {J_1(\nu_n \rho/a)}{J_1(\nu_n)}}
     \left [A_n e^{ih_n^a d}+A_{-n}e^{-ih_n^a d}\right ]+
     \beta E^a_\rho(\rho,\omega)e^{i\frac \omega v d}\quad\mbox{при}\quad 0<\rho<a\;.\]

      Домножая уравнение (1.8) на \(\rho\,J_1(\nu_m \rho /b)\)   и интегрируя по \(\rho\) от
    0 до \(b\), получаем бесконечную систему линейных алгебраических уравнений для искомых коэффициентов:

      \[h_n^b(B_ne^{ih_n^bd}-B_{-n}e^{-ih_n^bd})=
      \mu_n\sum_{s=1}^\infty(A_se^{ih_s^ad}-A_{-s}e^{-ih_s^ad})\frac {h_s^a}
      {\nu_s^2-(\nu_n a/b)^2}\;+\]
      \vspace{-0.7cm}\[\eqno(1.10)\]\vspace{-0.5cm}
      \[+\frac{2e^{i\frac \omega v d}}{bJ_1(\nu_n)}\Big[ \int_0^a{E_\rho^a(\rho,\omega)
      J_1(\nu_n\rho /b)\rho \,\mathrm{d}\rho}-\int_0^b{E_\rho^b(\rho,\omega)
      J_1(\nu_n\rho/ b)\rho \,\mathrm{d}\rho}\Big],\;n=1,2,\ldots\,,\]
        где
        \[\mu_n=2\Big(\frac a b\Big)^2\,\frac{\nu_n J_0(\nu_n a
        /b)}{J_1(\nu_n)}\,.\eqno(1.11)\]

    Домножая уравнение (1.9) на \(\rho\,J_1(\nu_m \rho /a)\)
     и интегрируя по \(\rho\) от 0 до \(a\), получаем ещё одно соотношение для этих же коэффициентов:

       \[A_ne^{ih^a_nd}+A_{-n}e^{-ih_n^ad}=\sum_{s=1}^{\infty}(B_se^{ih_s^bd}+
      B_{-s}e^{-ih_s^bd})\frac {\mu_s} {\nu^2_n-(\nu_s a/ b)^2}\,+\]
        \vspace{-0.8cm}\[\eqno(1.12)\]\vspace{-0.7cm}
       \[+\frac {2\beta e^{i\frac \omega v d}}{kaJ_1(\nu_n)} \int_0^a\big[E_\rho^b(\rho,\omega)
       -E_{\rho}^a(\rho,\omega)\big]J_1(\nu_n \rho /a)\rho\,\mathrm{d}\rho,\quad n=1,2,\ldots\,.\]

          Условие трансляционной симметрии (1.1) для Фурье-компонент поля имеет вид
       \[\left.\begin{array}{lll}\rv E(\rho,\,z=2D-d,\omega)&=&\rv E(\rho,\,z=-d,\omega)\,e^{i\psi}\,,\\[0.2cm]
       \rv H(\rho,\,z=2D-d,\omega)&=&\rv H(\rho,\,z=-d,\omega)\,e^{i\psi}\,,\end{array}\right\}
       \eqno(1.13)\]
       где \(\psi\) -- набег фазы на периоде структуры \(2D\), равный
       \vspace{-3mm}
       \[\psi=2D\frac \omega v\,,\eqno(1.14)\]
       и приводит к следующим соотношениям для коэффициентов разложения:

       \vspace{1mm}
       электрическое поле  ---
       \vspace{1mm}
       \[h_n^b(B_ne^{-ih_n^bd}-B_{-n}e^{ih_n^bd})=\]
       \[=\mu_ne^{-i\psi}
       \sum_{s=1}^\infty\Big[A_se^{ih_s^a(2D-d)}-A_{-s}e^{-ih_s^a(2D-d)}\Big]
       \frac {h_s^a}{\nu_s^2-(\nu_n a/b)^2}\,+\eqno(1.15)\]
        \[+\frac {2e^{-i\frac \omega v d}}{bJ_1(\nu_n)}\Big[
        \int_0^aE^a_\rho(\rho,\omega)J_1(\nu_n\rho/ b)\rho\,\mathrm{d}\rho-
        \int_0^bE^b_\rho(\rho,\omega)J_1(\nu_n\rho/ b)\rho\,\mathrm{d}\rho\Big],\]

       \vspace{1mm}
        магнитное поле  ---
         \vspace{1mm}
        \[A_ne^{ih_n^a(2D-d)}+A_{-n}e^{-ih_n^a(2D-d)}=\]
        \[=e^{i\psi}\sum_{s=1}^\infty\Big(B_se^{-ih_s^bd}+B_{-s}e^{ih_s^bd}\Big)
        \frac {\mu_s}{\nu_n^2-(\nu_s a/b)^2}\,+\eqno(1.16)\]
        \[ +\frac{2\beta e^{i(\psi -\frac \omega v d)}}{kaJ_1(\nu_n)}\int_0^a\big[
        E_\rho^b(\rho,\omega)-
        E_\rho^a(\rho,\omega)\big]J_1(\nu_n\rho/ a)\rho\,\mathrm{d}\rho\,.\]

         Вводя вспомогательные коэффициенты
       \[\left.\begin{array}{lll}D_n&=&(B_n+B_{-n})\cos h_n^bd\,,\\[0.3cm]
        E_n&=&(B_n-B_{-n})\sin h_n^bd\,,\\[0.3cm]
        F_n&=&(A_ne^{ih_n^aD}+A_{-n}e^{-ih_n^aD})\,e^{-i\psi/2}\cos h_n^al\,,\\[0.3cm]
        G_n&=&(A_ne^{ih_n^aD}-A_{-n}e^{-ih_n^aD})\,e^{-i\psi/2}\sin h_n^al\,,
        \end{array}\right\}\eqno(1.17)\]
        вычисляя интегралы
        \[\int_0^bE_\rho^b(\rho,\omega)J_1(\nu_n\rho/ b)\rho\,\mathrm{d}\rho-
        \int_0^aE_\rho^a(\rho,\omega)J_1(\nu_n\rho/ b)\rho\,\mathrm{d}\rho
        =\frac{Qb}{\pi v}\,\frac{\nu_nJ_0(\nu_n a/b)}
        {(\Gamma b)^2+\nu_n^2}\,\frac 1{I_0(\Gamma a)}\,,\]

        \[\int_0^a\big[E_\rho^b(\rho,\omega)-E^a_\rho(\rho,\omega)\big]
        J_1(\nu_n\rho/ a)\rho\,\mathrm{d}\rho=\frac{Qa}{\pi v}\,\frac{(\Gamma a)^2I_0(\Gamma a)
        J_1(\nu_n)}{(\Gamma a)^2+\nu_n^2}\,
        \Big[\frac{K_0(|\Gamma|b)}{I_0(\Gamma b)}
        -\frac{K_0(|\Gamma|a)}{I_0(\Gamma a)}\Big]\,,\]
        \noindent
        после несложных алгебраических преобразований
        получаем бесконечную систему линейных алгебраических уравнений
        относительно коэффициентов \(D_n, E_n\), которую удобно записать в матричном виде:
        \[\rv M \;\rv X=\rv F\;.\eqno(1.18)\]

        При численном решении бесконечной системы уравнений е\"е приходится редуцировать,
         то есть оставлять конечное число уравнений и неизвестных. Далее  подразумевается,
         что матрица   \(\rv M \) квадратная и состоит из четыр\"ех квадратных матричных
          блоков, имеющих размерность \(N\):

         \[\rv M=\left (\begin {array}{cc}\rv P\,\sin\psi /2&\rv Q\,\cos  \psi /2\\[0.5cm]
        \rv R\,\cos \psi /2&\rv S\,\sin \psi /2\end{array}\right )\;,\eqno(1.19)\]
        так что размерность матрицы \(\rv M \) равна \(2N\).

         Элементы блоков матрицы имеют следующий вид:
        \[\left.\begin{array}{lclcl}P_{mn}&\!=\!&\phantom{-}\delta_{mn}\,h_n^b a\,\tg h_n^b d&-&\beta_{mn}\,,\\[0.3cm]
        Q_{mn}&\!=\!&\phantom{-}\delta_{mn}\,h_n^b a\,\ctg h_n^b d&+&\beta_{mn}\,,\\[0.3cm]
        R_{mn}&\!=\!&\phantom{-}\delta_{mn}\,h_n^b a\,\tg h_n^b d&+&\alpha_{mn}\,,\\[0.3cm]
        S_{mn}&\!=\!&\!\!-\; \delta_{mn}\,h_n^b a\,\ctg h_n^b d&+&\alpha_{mn}\,,\end{array}
        \right \}\;\qquad m,n=1,2,\ldots,N\,,\eqno(1.20)\]
        где \(\delta_{mn} -\) символ Кронеккера,
        \[\alpha_{mn}=\mu_m\mu_n\sum_{i=1}^I\frac{h_i^a a\,\tg h_i^a l}
        {[\nu_i^2-(\nu_n a/ b)^2][\nu_i^2-(\nu_m a/ b)^2]}\,,\]
        \[\beta_{mn}=\mu_m\mu_n\sum_{i=1}^I\frac{h_i^a a\,\ctg h_i^a l}
        {[\nu_i^2-(\nu_n a/ b)^2][\nu_i^2-(\nu_m a/ b)^2]}\,,\]
         \(\mu_n\) определены в (1.11), а верхний предел суммирования \(I\) выбирается
        достаточно большим, во всяком случае \(I>N\).
    \noindent
     В том же порядке редукции размерность вектора-столбца решения уравнения (1.18)
     \(\rv X\) равна \(2N\): первые \(N\) элементов составляют коэффициенты \(D_n\),
     последующие -- \(E_n\).
         Вектор-столбец \(\rv F\) правой части уравнения (1.18), естественно, имеет ту же размерность
         \(2N\); его последовательные  элементы \(U_n\) и \(V_n\)  имеют следующий вид:
        \[\left.\begin{array}{lcl}U_n&=&\displaystyle{\frac {Q\mu_n}{\pi c \beta}}
        \big(-\xi_n\cos\displaystyle{\frac \omega v l}\,+\,\eta
        \sigma_{2n}\sin\displaystyle{\frac\omega v l}\big)
        = \frac{Qa}{\pi c }\;u_n\,,\\[0.5cm]
        V_n&=&\displaystyle{\frac {Q\mu_n}{\pi c \beta}}\big(\phantom{-}\xi_n\sin\displaystyle{\frac \omega v l}
        \,-\,\eta\sigma_{1n}\cos\displaystyle{\frac\omega v l}\big)=\frac{Qa}{\pi c}\;v_n\,,\end{array}\right \}\quad n=1,2,\ldots,N\,,\eqno(1.21)\]
         где
          \[\xi_n=\Big(\frac b a\Big)^2\frac 1{I_0(\Gamma a)
        [(\Gamma b)^2+\nu_n^2]}\,,\]

        \[\eta=\frac{2\Gamma a}{\gamma}\Big[\frac{K_0(|\Gamma| b)}
        {I_0(\Gamma b)}I_0(\Gamma a)-K_0(|\Gamma| a)\Big]\,,\]

        \[\sigma_{1n}=\sum_{i=1}^I\frac{h^a_i a\,\tg
        h_i^al}{\big[\nu_i^2-(\nu_n a/ b)^2\big]\big[(\Gamma
        a)^2+\nu_i^2\big]}\,,\]

        \[\sigma_{2n}=\sum_{i=1}^I\frac{h^a_i a\,\ctg
        h_i^al}{\big[\nu_i^2-(\nu_n a/ b)^2\big]\big[(\Gamma
        a)^2+\nu_i^2\big]}\,.\]

\vspace{2mm}
        Привед\"ем ещё формулы, с помощью которых по найденным в результате решения
         уравнения (1.18) коэффициентам \(D_n\,,\;E_n\) выражаются коэффициенты \(F_n\,,\;G_n\,,\)
        определяющие поля в области \(A\,:\)
         \[\left. \begin{array}{lcl}F_n&=&\displaystyle{\sum_{m=1}^N \Big( D_m\cos\frac\psi 2+
         E_m\sin\frac\psi 2\Big)\frac {\mu_m} {\nu_n^2-\!(\nu_m a/ b)^2}}+
         {\frac Q {\pi v}\,\frac {\eta} {(\Gamma a)^2+\nu_n^2}\,\cos \frac \omega v l} ,\\[0.5cm]
         G_n&=&\displaystyle{\sum_{m=1}^N\Big( D_m\sin\frac\psi 2-
         E_m\cos\frac\psi 2\Big)\frac {\mu_m} {\nu_n^2-\!(\nu_m  a/ b)^2}}+
         {\frac Q {\pi v}\,\frac {\eta} {(\Gamma a)^2+\!\nu_n^2}\,\sin\frac \omega v l} .
         \end{array}\right\}\eqno(1.22)\]
        Отметим, что при выбранном представлении (1.17) коэффициентов
        \(D_n,\,E_n,\,F_n,\,G_n\)  все элементы матрицы  \(\rv M\) (1.19) -- 
        действительные безразмерные  величины  во всей области
        частот \(\omega\), в том числе и там, где продольные  волновые числа круглого волновода
        \(h_n^R\) чисто мнимые; при этом блочные матрицы \(\rv P, \rv Q, \rv R, \rv S\) -- симметричные.

        Совокупность формул (1.1)\,--\,(1.22) полностью определяет Фурье-компоненты всех
        составляющих поля заряда, движущегося вдоль оси КДВ.
        Однако такое представление носит в значительной степени формальный характер
        и мало о ч\"ем говорит  до тех пор, пока не определена методика решения уравнения (1.18) и
        не выявлены особенности вычисления интегралов Фурье (1.3).

        Формальное решение уравнения (1.18) легко представить с помощью обратной матрицы:
        \[\rv X = \rv M^{-1}\rv F\,.\eqno(1.23)\]
Однако такой вид решения матричного уравнения теряет смысл на тех частотах \(\omega\),  на
        которых определитель матрицы \(|\rv M|=0\).  При определ\"енных условиях в этом случае система линейных алгебраических уравнений может иметь отличные от нуля решения и в том случае, когда она однородная, то есть когда при е\"е матричной записи (1.18) вектор-столбец  \( \rv F=0\). Такие решения
        определяют собственные волны КДВ, прич\"ем для замкнутых периодических структур они хорошо изучены. Однако нам необходимо привести здесь конкретные выражения для полей и энергетических
        характеристик этих волн при выбранном нами способе представления результатов. Как выяснится из дальнейшего, именно эти волны определяют потери энергии заряда на излучение при движении в КДВ .

        %\newpage
\vspace{1.0cm}
          \begin{center}
   \large\textbf{2. Собственные волны круглого диафрагмированного волновода}
   \end{center}

    При фиксированных геометрических размерах структуры элементы
    матрицы \(\rv M\) являются функциями двух переменных: набега фазы волны на периоде структуры
    \(\psi\) и  частоты \(\omega\). В  задаче об излучении равномерно движущегося заряда переменные
    \(\psi\) и \(\omega\)  связаны соотношением (1.14), вытекающим из трансляционной симметрии задачи;
     при анализе же свойств  собственных волн периодической структуры их следует рассматривать как независимые переменные.

    Каждому значению  \(\psi\)  соответствует бесконечный дискретный спектр значений
    \(\omega_s(\psi)\), при которых обращается в нуль определитель  матрицы \(\rv M\).
    Из анализа структуры блоков матрицы \(\rv M\)следует,  что  функция \(\omega_s(\psi) \), представляющая
    собой дисперсионную кривую \(s\)-й собственной волны, является   ч\"етной и  периодической, так что
    достаточно знать е\"е поведение в  интервале \(0\leqslant\psi\leqslant\pi\).

    При \(\psi=0\) нули определителя матрицы \(\rv M\)
     совпадают с нулями определителя одной из матриц \(\rv Q\) или \(\rv R\), элементы которых определены
     формулами (1.18), а при \(\psi=\pi\)  с нулями
    определителя матрицы \(\rv P\)  или \(\rv S\).  Значения \(\omega_s(0)\) и  \(\omega_s(\pi)\) имеют
     определ\"енный физический смысл.    Они являются собственными частотами вспомогательного резонатора,
      представляющего собой половину рассматриваемой ячейки КДВ с торцевыми стенками при \(z=0\) и \(z=D\).
      Нули определителя  матрицы \(\rv Q\) определяют  собственные частоты резонатора,
       у которого обе торцевые стенки идеально проводящие, нули определителя матрицы \(\rv R\)  --
       собственные частоты резонатора с <<идеальными магнитными>>  торцевыми стенками
       (на таких стенках равна нулю составляющая магнитного поля \(H_{\varphi}\) и нормальная
       составляющая электрического поля \(E_z\)).
       Совокупность собственных частот  этих резонаторов на плоскости переменных \((\psi,\omega )\)
       определяет  точки пересечения  всех дисперсионных кривых \(\omega_s(\psi) \) собственных волн КДВ
       с линией \(\psi = 0\).  Совокупность собственных частот резонаторов, у которых одна из стенок
       идеально проводящая, другая -- <<идеальная магнитная>>, определяют точки пересечения дисперсионных кривых КДВ с линией \(\psi=\pi\). Нули определителя матрицы \(\rv P\) соответствуют собственным частотам  резонатора, у которого идеально проводящей является стенка \(z=0\); нули определителя матрицы \(\rv S\) ---   резонатора с идеально проводящей стенкой  \(z=D\).

     Согласно общей теории периодических структур [4] групповые скорости их собственных волн
     пропорциональны производной соответствующей дисперсионной кривой  и в КДВ  с
     пространственным периодом \(2D\) равны
     \[v_{gr}(s,\psi)=2D\frac {d\omega_s}{d\psi}\,.\eqno(2.1)\]
     При \(\psi=0\) и \(\psi=\pi\) поле собственной волны КДВ совпадает с полем собственного колебания
     соответствующего вспомогательного резонатора, которое представляет собой стоячую волну, следовательно, и групповая скорость,  и производные  \(d\omega_s/d\psi\)  в этих  точках равны нулю.

      Принцип причинности, требующий  выполнения неравенства \(v_{gr}<c\), накладывает
      ограничение на значение  производной
      \[\Big|\frac{d\omega_s}{d \psi}\Big|<\frac c{2D}\eqno(2.2)\]
      при всех значениях \(\psi\). Поэтому функция \(\omega_s(\psi) \) ограничена, а ширина полосы частот,
       в пределах которой может распространяться каждая собственная волна КДВ, всегда меньше \(\pi c/2D\).

       Из общих физических соображений следует, что последовательные дисперсионные кривые
       \(\omega_s(\psi)\) и  \(\omega_{s+1}(\psi)\)  не могут пересекаться. Функция \(\omega_s(\psi)\) на интервале \(\psi\) от 0 до \(\pi\) может быть монотонной, и тогда групповая скорость собственной волны на вс\"ем этом интервале имеет один и тот же знак.  Если при этом \(\omega_s(0)<\omega_{s+1}(\pi)\) и \(\omega_s(\pi)<\omega_{s+1}(0)\), то между полосами частот, в пределах  которых распространяются эти волны, имеется полоса частот, где распространяющихся волн нет; если же одно из этих условий не
       выполнено,  то полосы непрозрачности между этими собственными  волнами КДВ нет.

      Если же  хотя бы одна из соседних дисперсионных кривых  не является монотонной на интервале
      \(\psi\) от 0 до \(\pi\), то у неё на этом интервале обязательно должен быть по крайней мере один
      локальный  максимум или минимум. Каждый локальный экстремум соответствует стоячей волне в
       КДВ.   При условии, что экстремум  лежит вне интервала \([\omega_s(0), \omega_s(\pi)]\), полосы непрозрачности может не быть даже при выполнении  привед\"енных выше неравенств.

    Пространственная структура поля \(s\)-й собственной волны определяется
    вектором решения  однородной системы уравнений
    \[\rv M(\omega_s)\tilde{\rv  X}_s=0\,,\eqno(2.3)\]
     который в отличие   от решения  \(\rv X\) неоднородной системы (1.18)
      помечаем индексом  \(s\) и верхним значком  \(\tilde{\phantom{a}}\).

     Система уравнений (2.3)  является бесконечной, и поэтому сформулировать условие наличия решения
     у такой однородной системы  непросто.  При численном решении системы   матрицу \(\rv M\) необходимо  редуцировать  (большинство приводимых далее результатов получены для матрицы размерности \(N=500\)).

    Для квадратной матрицы \(\rv M\)  размерности \(N\),  согласно общей теории однородных
    линейных алгебраических уравнений [9], уравнение (2.3) может  иметь отличный от нуля вектор-столбец решения  \(\tilde{\rv  X}_s\) при выполнении двух условий: 1) \(|\rv M|=0\) и 2) ранг матрицы \(R<N\).

      Такой вектор-столбец будет единственным при условии \(N-R=1\), а из вида уравнения (2.3) очевидно,
      что он определяется с точностью до произвольного комплексного множителя. Эта неопредел\"енность решения однородной системы  отражает  произвольность амплитуды \(\mathcal{A}_s\) и
      фазы \(\theta_s\)  собственной волны.

        Вектор решения \(\tilde{\rv  X}_s\)  может быть найден на основе известного свойства  вырожденной
     матрицы: все столбцы соответствующей   присоедин\"енной матрицы линейно зависимы, и любой из
    них может  быть взят в качестве решения  однородной системы (вырожденной   называется матрица, определитель которой  равен нулю, а присоедин\"енной --- транспонированная матрица алгебраических
    дополнений соответствующих элементов).

  Однако такой строго обоснованный способ решения однородной системы линейных алгебраических уравнений при численной его реализации для рассматриваемой конкретной задачи не удобен. Нами был использован другой способ, строгое математическое обоснование которого здесь опускается, но приводится его 
  краткое  описание.

  Пусть имеется матрица, элементы которой являются функциями частоты \(\omega\), такая, что на некоторых частотах \(\omega_s\) определитель этой матрицы имеет нули первого порядка, то есть $|\rv M(\omega)| \to (\omega-\omega_s)\,C$ при $\omega \to \omega_s $,  
  где \(C\neq 0\). На частоте \(\omega_s\) обратная матрица \(\rv M^{-1}\) не существует, но существует предел
  \[\tilde{\rv M}(\omega_s)=\lim_{\omega \to \omega_s}\big[\frac{\omega-\omega_s}{\omega_s}\,
  \rv M^{-1}(\omega)\big]\,.\eqno(2.4)\]

  У матрицы \(\tilde{\rv M}(\omega_s)\) все столбцы и все строки линейно зависимы; все элементы матрицы \(\tilde{m}_{i,n}\) связаны соотношением
   \[\tilde{m}_{m,n}=\frac {\tilde{m}_{i,n}\cdot\tilde{m}_{m,i}}{\tilde{m}_{i,i}}\,,\quad i=1,2,\ldots\,,\]
   то есть матрица полностью определяется  двумя векторами. В дальнейшем в качестве таких, однозначно определ\"енных векторов, будем использовать 1-й столбец и 1-ю строку матрицы
   \(\tilde{\rv M}(\omega_s)\).

  Любой столбец матрицы \(\tilde{M}(\omega_s)\)  представляет собой  решение  матричного уравнения (2.3). В дальнейшем под компонентами вектора \(\tilde{\rv X}_s(\omega_s)\), то есть коэффициентами \(\tilde{D}_{s,n},\tilde{E}_{s,n}\), будем понимать 1-й столбец матрицы \(\tilde{M}(\omega_s)\),  а неопредел\"енность решения уравнения (2.3), выраженную в виде произвольного комплексного множителя, будем учитывать в амплитуде   \(\mathcal{A}_s\) и фазе  \(\theta_s\)  собственной волны КДВ.

   В результате в соответствии с формулами (1.3) и (1.7) при \(\rv E^R(\rv r, \omega)=0\)
  составляющие поля \(s\)-й  собственной волны КДВ в области \(B\)  ячейки можно записать в
   следующем действительном виде:

   \[\left.\begin{array}{lcl}H_{s,\varphi}(\rv r,t)&=&
    \displaystyle{
    \frac{ \mathcal{A}_s k_s}{b}\sum_{n=1}^N\frac{J_1(\nu_n\rho/b)}
    {J_1(\nu_n)}}\times\\[0.5cm] &&\times\displaystyle{\Big[\tilde{D}_{s,n}\,\frac{\cos h_n^bz}{\cos
    h_n^bd}\,\cos(\omega_s t\!+\!\theta_s)+\tilde{E}_{s,n}\,\frac{\sin h_n^bz}{\sin
    h_n^bd}\,\sin(\omega_s t\!+\!\theta_s)\Big]},\\[0.6cm]
    E_{s,\rho}(\rv r,t)&=&
    \displaystyle{\frac {\mathcal{A}_s} b\sum_{n=1}^N\frac{J_1(\nu_n\rho/b)}
    {J_1(\nu_n)}\,h_n^b}\times\\[0.5cm]&&\times\displaystyle{\Big[\tilde{D}_{s,n}
    \,\frac{\sin h_n^bz}{\cos h_n^bd}\,\sin(\omega_s t\!+\!\theta_s)+\tilde{E}_{s,n}\,\frac{\cos h_n^bz}
    {\sin h_n^bd}\,\cos(\omega_s t\!+\!\theta_s)\Big]}\,,\\[0.6cm]
    E_{s,z}(\rv r,t)&=&
    \displaystyle{\frac  {\mathcal{A}_s} {b^2}\sum_{n=1}^N\frac{\nu_n  J_0(\nu_n\rho/b)}
    {J_1(\nu_n)}}\times\\[0.5cm]&&\times\displaystyle{\Big[\tilde{D}_{s,n}\,\frac{\cos h_n^bz}{\cos
    h_n^bd}\,\sin(\omega_s t\!+\!\theta_s)-\tilde{E}_{s,n}\,\frac{\sin h_n^bz}{\sin
    h_n^bd}\,\cos(\omega_s t\!+\!\theta_s)\Big]}\,.\end{array}\right\}\eqno(2.5)\]

Составляющие поля  в области A определяются коэффициентами \(\tilde{F}_{s,n}\,,\tilde{G}_{s,n}\),
которые  связаны с   \(\tilde{D}_{s,n},\;\tilde{E}_{s,n}\) соотношениями (см. также (1.22))
  \[\left. \begin{array}{lcl}\tilde{F}_{s,n}&=&\displaystyle{\sum_{m=1}^N
  \Big( \tilde{D}_{s,m}\cos{\frac\psi  2 }+\tilde{E}_{s,m}\sin {\frac\psi 2}\Big)\frac {\mu_m} {\nu_n^2-\!(\nu_m a/b)^2}}\,,\\[0.5cm]
        \tilde{G}_{s,n}&=&\displaystyle{\sum_{m=1}^N
  \Big( \tilde{D}_{s,m}\sin{\frac\psi 2} - \tilde{E}_{s,m}\cos {\frac\psi 2}\Big)\frac {\mu_m} {\nu_n^2-\!(\nu_m a/b)^2}}\,,
         \end{array}\right\}\quad n=1,2,\ldots,N\,,\eqno(2.6)\]
и могут быть записаны в виде
       \[\left.\begin{array}{l}\phantom{aaaaaaaaaaaaaa}H_{s,\varphi}(\rv r,t)=
    \displaystyle{
    \frac{\mathcal{A}_ s k_s}{a}\sum_{n=1}^N\frac{J_1(\nu_n\rho/a)}
    {J_1(\nu_n)}}\;\times\\[0.5cm] \times\displaystyle{\Big[\tilde{F}_{s,n}\,\frac{\cos h_n^a (z\!-\!D)}{\cos
    h_n^a l}\,\cos(\omega_s t\!-\!\frac \psi 2\!+\!\theta_s)+\tilde{G}_{s,n}\,\frac{\sin h_n^a (z\!-\!D)}{\sin
    h_n^a l}\,\sin(\omega_s t\!-\!\frac \psi 2 \!+\!\theta_s)\Big]},\\[0.6cm]
    \phantom{aaaaaaaaaaaaaa}E_{s,\rho}(\rv r,t) =
    \displaystyle{\frac {\mathcal{A}_ s} a\sum_{n=1}^N\frac{J_1(\nu_n\rho/a)}
    {J_1(\nu_n)}\,h_n^a}\;\times\\[0.5cm]\times\displaystyle{\Big[\tilde{F}_{s,n}
    \,\frac{\sin h_n^a (z\!-\!D)}{\cos h_n^a l}\,\sin(\omega_s t\! -\!\frac \psi 2 \!+\!\theta_s)
    +\tilde{G}_{s,n}\,\frac{\cos h_n^a (z\!-\!D)}
    {\sin h_n^a l}\,\cos(\omega_s t\!-\!\frac\psi 2 \!+\!\theta_s)\Big]}\,,\\[0.6cm]
    \phantom{aaaaaaaaaaaaaa}E_{s,z}(\rv r,t) =
    \displaystyle{\frac {\mathcal{A}_ s} {a^2}\sum_{n=1}^N\frac{\nu_n  J_0(\nu_n\rho/a)}
    {J_1(\nu_n)}}\;\times\\[0.5cm]\times\displaystyle{\Big[\tilde{F}_{s,n}\,\frac{\cos h_n^a(z\!-\!D)}{\cos
    h_n^a l}\,\sin(\omega_s t\!-\!\frac \psi 2 \!+\!\theta_s)-\tilde{G}_{s,n}\,\frac{\sin h_n^a (z\!-\!D)}{\sin
    h_n^a l}\,\cos(\omega_s t\!-\!\frac \psi 2\!+\!\theta_s )\Big]}\,.\end{array}\right\}\eqno(2.7)\]

    Для упрощения записи этих (и последующих) формул у величин \(h_n^b\) и \(h_n^a\)
  опущен индекс \(s\), соответствующий номеру собственной волны,
  хотя в формулах, определяющих эти величины, следует  заменить \(k\) на \(k_s\) (смотри также (1.1)).

Поле в любой точке КДВ, отстоящей от соответствующей точки \(z\) рассматриваемой
   ячейки на \(n\)  периодов, определяется согласно трансляционной симметрии:
    \[\rv E_s(\rho,z+2Dn, t+n\psi/\omega_s)=\rv E_s(\rho,z,t)\,.\]

Направление распространения собственной волны КДВ определяется знаком \(\psi\).
   Из структуры матрицы \(\rv M\)  ясно, что \(\omega_s\) и \(\tilde{D}_{s,n}\) ч\"етные функции \(\psi\),
   а \(\tilde{E}_{s,n}\) -- неч\"етные; из (2.5) следует, что \(\tilde{F}_{s,n}\) ч\"етные функции \(\psi\),
   а \(\tilde{G}_{s,n}\) -- неч\"етные. Поэтому для составляющих поля собственной волны КДВ выполняются соотношения, аналогичные соотношениям для собственной волны однородного круглого волновода:
    \[ \left.\begin{array}{lcl}E_z(-\psi,-z)&=&E_z(\psi,z)\,,\\[0.2cm]
    E_{\rho}(-\psi,-z)&=&-E_{\rho}(\psi,z)\,,\\[0.2cm]
    H_{\varphi}(-\psi,-z)&=&H_{\varphi}(\psi,z)\,.\end{array}\right\}\]

        Основными энергетическими характеристиками собственной волны
    КДВ являются усредн\"енные по периоду колебания потоки вектора
    Пойнтинга через поперечное сечение волновода

    \[\overline{P}_s(z)=\frac c {4\pi}\overline{\int\limits_S E_{s,\rho}(\rv r,t)\,H_{s,\varphi}(\rv
    r,t)\,\mathrm{d}S}\]
    и приходящаяся на один период структуры энергия поля
    \[\overline{W}_s=\overline{W}_s^H+\overline{W}_s^E=\frac 1 {8\pi}\overline{ \int\limits_VH_{s,\varphi}^2(\rv
    r,t)\,\mathrm{d}V}
    +\frac 1 {8\pi}\overline{ \int\limits_V\big[E_{s,\rho}^2(\rv r,t)+E_{s,z}^2(\rv r,t)\big]\,\mathrm{d}V}
    \,,\]
    которые с помощью формул (2.5) и (2.7) выражаются в следующем
    виде:

    поток в области \(A\) равен
    \[\overline{P}_s^A=\frac {\mathcal{A}^2_s\omega_s} 4 \sum_{n=1}^N\tilde{F}_{s,n}\tilde{G}_{s,n}\frac{h_n^a}{\sin
    2h_n^a l}\;,\eqno(2.8)\]

    поток в области \(B\) равен
    \[\overline{P}_s^B=\frac {\mathcal{A}^2_s\omega_s} 4 \sum_{n=1}^N\tilde{D}_{s,n}\tilde{E}_{s,n}\frac{h_n^b}{\sin
    2h_n^b d}\;,\eqno(2.9)\]

    энергия магнитного поля определяется выражением
    \[\begin{array}{l}\overline{W}_s^H=\displaystyle{\frac{\mathcal{A}^2_s k_s^2}{16}
    \sum_{n=1}^N\Big\{l\Big[\tilde{F}_{s,n}^2\big(1\!+\!
    \tg^2h_n^a l\!+\!\frac{\tg h_n^a l}{h_n^a l}\big)+
    \tilde{G}_{s,n}^2\big(1\!+\!
    \ctg^2h_n^a l\!-\!\frac{\ctg
    h_n^a l}{h_n^a l}\big)\Big]}+\\[0.7cm]
    \phantom{W_saaaaaaaa}+\,\displaystyle{
    d\Big[\tilde{D}_{s,n}^2\big(1\!+\!\tg^2h_n^b d
    \!+\!\frac{\tg h_n^b d}{h_n^b d}\big)+
    \tilde{E}_{s,n}^2\big(1\!+\!
    \ctg^2 h_n^b d\!-\!\frac{\ctg h_n^b d}{h_n^b d}\big)\Big]\Big\}}\,,
    \end{array}\eqno(2.10)\]

    а энергия электрического поля равна
\[\overline{W}_s^E=\overline{W}_s^H\!-\displaystyle{\frac {\mathcal{A}^2_s} 8\sum_{n=1}^N\Big[h_n^b
    \big(\tilde{D}_{s,n}^2\tg h_n^b d\!-\!\tilde{E}_{s,n}^2\ctg h_n^b d\big)}
   \!+h_n^a\big(\tilde{F}_{s,n}^2\tg h_n^a l-\tilde{G}_{s,n}^2\ctg h_n^a l
    \big)\Big].\eqno(2. 11)\]

Общая теория собственных волн в периодических структурах к настоящему времени разработана очень
основательно [4].  В частности, согласно теоремам Белла должно выполняться равенство средних по периоду колебания магнитной и электрической энергии, приходящихся на пространственный период структуры.
Поэтому среднее значение   энергии электромагнитного поля \(s\)-й волны в ячейке \[\overline{W}_s=2\overline{W}_s^H.\eqno(2.12)\]

  Ещ\"е одним общим для всех периодических структур свойством является связь между средней энергией
    поля, приходящейся на пространственный период структуры, средним потоком вектора Пойнтинга и
    групповой скоростью волны:
    \[\overline{P}_s= v_{gr}\,\frac {\overline{W}_s}{2D} \,.\eqno(2.13)\]
    Это соотношение, наравне с (2.1), может служить ещ\"е одним определением групповой скорости волны.
   Очевидно также, что средние потоки вектора Пойнтинга в областях  А и В ячейки, определяемые формулами (2.8) и (2.9), должны быть равны.

   Доказательство этих свойств аналитическими методами на основании привед\"енных выше формул
  довольно громоздко и затруднительно. Но, во-первых, в этом нет необходимости, и, во-вторых, целесообразнее   убедиться в выполнении этих соотношений при отладке программы вычислений на компьютере и использовать их для контроля достигнутой точности в каждом варианте расч\"етов.

  %   \newpage
    \vspace{1.0cm}
    \begin{center}
   \large\textbf{3. Временная зависимость поля, возбуждаемого  \\в ячейке зарядом}
   \end{center}

   Перейд\"ем теперь к анализу поведения поля, возбуждаемого в ячейке движущимся зарядом,
   в функции времени.  Привед\"енные в разделе \textbf{1} формулы позволяют записать продольную
   составляющую электрического поля в виде суммы интегралов:

   в области \(B\)
    \[E_z(\rv r,t)\!=\!E_z^b(\rv r,t) \!+\!\frac 1{b^2}\sum_{n=1}^{\infty}\frac{\nu_n J_0(\nu_n\rho/b)}{J_1(\nu_n)}
    \!\int\limits_{-\infty}^\infty\! e^{-i\omega t}
    \,\Big(iD_n\frac{\cos h_n^b z}{\cos h_n^b d}-E_n\frac{\sin h_n^b z}{\sin h_n^b d}\Big)\,\mathrm{d}\omega
    \,,\eqno(3.1)\]

   в области \(A\)
    \[\left.\begin{array}{lcl}E_z(\rv r,t)&=&E_z^a(\rv r,t)+\displaystyle{\frac 1{a^2}\sum_{n=1}^{\infty}\frac{\nu_n J_0(\nu_n\rho/a)}
    {J_1(\nu_n)}}\;\times\\[0.5cm]
    &\times&\displaystyle{\int\limits_{-\infty}^\infty e^{i(\psi/2-\omega t)}
    \Big[iF_n\frac{\cos h_n^a(D\!-\!z)}{\cos h_n^al}+G_n\frac{\sin h_n^a(D\!-\!z)}{\sin h_n^al}\Big]\,
    \mathrm{d}\omega}\, ,\end{array}\right\}\eqno(3.2)\]
   где \(E_z^b(\rv r,t)\) и \(E_z^a(\rv r,t)\) определены в (1.2).
   Аналогичным образом записываются составляющие поля  \(E_\rho(\rv r,t)\) и \(H_\varphi(\rv r,t)\). Отметим, что   составляющая поля \(E_z(\rv r,t)\) представляет особый интерес, поскольку только она отлична от нуля вдоль траектории заряда  и непосредственно взаимодействует с ним.

  Функцией  \(\omega\) в подынтегральных выражениях формул (3.1) и (3.2) являются волновые числа \(h_n^a\) и \(h_n^b\),  а также коэффициенты \(D_n,\;E_n,\;F_n,\;G_n,\) определяемые из системы уравнений (1.18). Отметим, что коэффициенты   \(D_n\) и \(F_n\) --  неч\"етные функции \(\omega\), а \(E_n\) и \(G_n\) -- ч\"етные, и поэтому оба интеграла действительные
  функции \(z\) и \(t\).

  Формулы (3.1) и (3.2) были получены в предположении идеальной проводимости стенок КДВ. Принцип причинности требует,   чтобы стенки обладали поглощением, хотя бы и предельно малым. Для рассмотренных в разделе \textbf{2} собственных волн такое   поглощение несущественно: оно не меняет структуры поля во вс\"ем объ\"еме КДВ и практически не сказывается на характере их
  распространения. Поглощение приводит только к появлению бесконечно малой мнимой добавки к собственной частоте волны,   что обеспечивает е\"е затухание через достаточно большой промежуток времени.

  Для расч\"ета излучения движущегося заряда в периодической структуре наличие поглощения имеет принципиальное значение:   без его уч\"ета формулы (3.1) и (3.2) являются неопредел\"енными. Эта неопредел\"енность обусловлена тем, что на тех частотах, где определитель матрицы \(\rv M\) в уравнении (1.18) обращается в нуль, решение для вектора \(\rv X\), и, следовательно, для всех коэффициентов \(D_n,\;E_n,\;F_n,\;G_n\) обращается в бесконечность. Интегралы в (3.1), (3.2) становятся несобственными и необходимо установить их сходимость. Отметим, что из структуры элементов матрицы \(\rv M\) (1.19) следует, что частоты, на которых обращаются в нуль функции \(\cos h_n^b d,\,\sin h_n^b d,\,\cos h_n^al,\,\sin h_n^al\), не являются особенностями подынтегральных выражений.

  Положение нулей определителя матрицы \(\rv M\) на плоскости переменных \((\psi\,,\omega)\) определяется пересечением  дисперсионных кривых собственных волн КДВ с прямой <<линией заряда>>, описываемой формулой (1.14). Рассмотрим поведение коэффициентов \(D_n,\;E_n,\;F_n,\;G_n\) в ближайшей окрестности
  этих частот \(\omega_s\). Решение  (1.23) матричного уравнения (1.18) можно записать в виде
   \[\rv X(\omega)=\frac {\omega_s}{\omega-\omega_s}\,\tilde{\rv M}(\omega_s)\rv F(\omega_s)\,.\]
  В соответствии с принятыми в разделе \textbf{2} обозначениями элементы матрицы
  \(\tilde{\rv M}(\omega_s)\) могут быть представлены в виде

  \[\tilde{m}_{ni}=\left \{
  \begin{array}{l}\displaystyle{\frac{\tilde{D}_{s,n}}{\tilde{D}_{s,1}}\,b_i }\qquad n=1,2,\ldots,N\,,\\[0.5cm]
   \displaystyle{\frac{\tilde{E}_{s,n-N}}{\tilde{D}_{s,1}}\,b_i }\qquad n=N+1,N+2,\ldots,2N\,,\end{array}
   \right.\]
   где \(b_i=\tilde{m}_{1i}\) -- элементы 1-й строки матрицы \(\tilde{\rv M}\).

   В результате все коэффициенты вблизи точек \(\omega=\omega_s\) ведут себя как
   \[D_n(\omega)=\frac{Qa}{\pi c}\,\frac{\omega_s}{\omega-\omega_s}\,\frac{\tilde{D}_{s,n}}{\tilde{D}_{s,1}}
   \sum_{i=1}^{N}(b_iu_i+b_{N+i}v_i)\,,\qquad n=1,2,\ldots,N\,,\eqno(3.3)\]
   где  \(u_i\) и \(v_i\)  определены в (1.21); поэтому несобственные интегралы (3.1) и (3.2) вдоль
   действительной оси \(\omega\)  расходятся.

   Уч\"ет бесконечно малой отрицательной мнимой добавки к \(\omega_s\), обеспечивающей затухание волны при \(t \to \infty\), приводит к смещению  особенностей коэффициентов  \(D_n,\;E_n,\;F_n,\;G_n\) в нижнюю полуплоскость комплексной переменной \(\omega\). Это делает значения интегралов вдоль действительной оси \(\omega\) определ\"енными и позволяет сравнительно просто вычислить их  в двух предельных случаях: \(t\to \infty\) и \(t\to - \infty\). Для этого целесообразно перейти к интегрированию в комплексной плоскости переменной \(\omega\) и использовать теорию вычетов.

      Возможность вычисления интеграла в этих двух предельных случаях  обусловлена тем, что множитель \(e^{-i\omega t}\) в подынтегральных выражениях (3.1) и (3.2) является
      преобладающим по сравнению с другими зависящими от \(\omega\) экспонентами,
      через которые выражаются функции типа \(\cos h_n^b d\,,\sin h_n^b d \). Это позволяет провести
      интегрирование по \(\omega\) пут\"ем  замыкания контура  интегрирования в комплексной плоскости дугой большого радиуса. Для возможности пренебречь вкладом интеграла  вдоль этой дуги необходимо замыкать контур при отрицательных значениях \(t\) в верхней полуплоскости,  а при положительных
       --- в нижней.

       Выявленные формулой (3.3) особенности коэффициентов \(D_n(\omega)\) на действительной оси \(\omega\) свидетельствуют о том, что все особенности являются полюсами первого порядка. Учитывая поведение
       элементов матрицы \(\rv M\) и компонент вектора-столбца \(\rv F\) при отрицательных значениях \(\omega\), нетрудно убедиться, что полюсами являются и значения \(\omega=-\omega_s\).

       Сумма вычетов подынтегрального выражения \(f(\omega)\) в (3.1) в полюсах   \(\omega=\omega_s\) и
       \(\omega=-\omega_s\) легко вычисляется при подстановке (3.3) и аналогичного выражения для \(E_n(\omega)\) в (3.1):
       \[\begin{array}{l} Res[f(\omega);\omega_s]+Res[f(\omega);-\omega_s]=-\displaystyle{\frac 1{2\pi i}}
        \;\mathcal{A}_s\times\\[0.7cm]
       \times \displaystyle{\Big[\tilde{D}_{s,n}\,\frac{\cos h_n^bz}{\cos h_n^b d}\,
       \cos \omega_s t+\tilde{E}_{s,n}\,\frac{\sin h_n^b z}{\sin h_n^b d}\,
       \sin\omega_s t\Big]}\,,\end{array}\eqno(3.4)\]
       где
        \[\mathcal{A}_s=\frac{4Qk_s a}{\tilde{D}_{s,1}}\sum_{i=1}^N(b_iu_i+b_{N+i}v_i)\,;\eqno(3.5)\]
ниже будет показано, что при больших  \(t\) параметр $\mathcal{A}_s$ соответствует амплитудам собственных волн в (2.5), (2.7).

        На комплексной плоскости \(\omega\), как в верхней, так в нижней полуплоскости, имеется ещ\"е бесконечное множество полюсов, расположенных на мнимой оси, и обусловленных, в частности, компонентами вектора-столбца \(\rv F \). Но вычеты в этих полюсах в обоих пределах \(t\to\infty\) и
        \(t\to -\infty\) стремятся к нулю. В результате при больших отрицательных значениях \(t\) (заряд ещ\"е далеко от рассматриваемой ячейки) поле очень мало и не имеет волнового характера. По своей структуре оно мало отличается от поля заряда в регулярном круглом волноводе (1.2), то есть экспоненциально спадает с расстоянием от точки нахождения заряда в данный момент времени.

        При больших положительных значениях \(t\) (заряд давно пролетел ячейку) заметный вклад в интеграл
        дают вычисленные выше вычеты (3.4) в полюсах, расположенных в непосредственной близости от действительной оси. Подставив значение интеграла, равное \(2\pi i\times\sum_{s=1}^\infty Res[f(\omega);\omega_s]+Res[f(\omega);-\omega_s],\) в (3.1) и сопоставив полученное выражение с составляющей \(E_z\) поля собственной волны КДВ (2.5), убеждаемся, что при больших \(t\) поле в ячейке асимптотически стремится к сумме собственных волн КДВ с амплитудами \(\mathcal{A}_s\), определяемыми формулой (3.5), и одинаковыми фазами \(\theta_s=\pi/2\). В области А ячейки все составляющие возбужд\"енных волн определяются формулами (2.6) и (2.7) с набегом фазы на периоде \(\psi=2D\omega_s/v\).

               Средний по времени поток вектора Пойнтинга через поперечное сечение КДВ определяется
               формулой (2.9) с найденным выше значением амплитуды (3.5), а среднее значение   энергии электромагнитного поля \(s\)-й волны в ячейке \(\overline{W}_s=2\overline{W}_s^H,\) где
            \(\overline{W}_s^H\) определено в (2.10).

               Отметим, что поле в ячейке  имеет трансляционный характер, но не является периодическим. Среди
      возбужд\"енных собственных волн имеются волны как с положительной, так и отрицательной
      групповой скоростью, прич\"ем в спектре частот эти волны в зависимости от скорости заряда располагаются  в разной последовательности.

        Анализ зависимости вектора \(\rv F\) от скорости заряда \(v\) позволяет сделать вполне определ\"енные заключения об     амплитудах возбуждаемых в ячейке собственных волн в двух предельных случаях:
        \(v\to 0\) и \(v\to c\).
     При малых скоростях \(v\) параметр \(\Gamma  =\omega /v\gamma\to\,\infty\). В этом пределе асимптотическое поведение функций  \(I_0\,, K_0\) приводит к экспоненциально малому значению амплитуд всех возбуждаемых собственных волн, так как в (1.21)
    \[\xi_n\to  \sqrt{2\pi}\big(\frac \omega v \,a\big)^{-3/2} e^{-\frac \omega v a},\quad
    \eta \to -\sqrt{ 2\pi\frac \omega v \,a} \,e^{-\frac \omega v a}\,,\]
   а нижняя частота спектра   собственных волн КДВ отлична от нуля, и для не\"е параметр \(ka\)  близок к 0,8.

   В другом предельном случае ультрарелятивистских скоростей \(\beta\to 1\) параметр \(\Gamma  \to 0\), так как    \(\gamma \to \infty\) и, следовательно,
    \[\eta \to 0\,,\quad\xi_n \to \Big(\frac b{a\nu_n}\Big)^2\,,\]
   то есть вектор-столбец \(\rv F\) переста\"ет зависеть  от энергии заряда. Поскольку при этом частоты возбуждаемых  собственных волн    стремятся к своему предельному значению, соответствующему набегу
    фазы на периоде \(\psi=2D k_s\), то можно говорить о  предельном спектре излучения.  Амплитуда
    каждой возбужд\"енной  собственной волны монотонно стремится к своему предельному значению с
    ростом  энергии заряда, но это предельное значение убывает с номером собственной волны.

    Из привед\"енных выше соотношений также очевидно, что при заданной скорости заряда амплитуды возбуждаемых  собственных волн начинают экспоненциально спадать, начиная с некоторого номера
    волны, тем большего, чем выше энергия заряда.

      Следует, однако, оговориться, что все рассуждения о предельном спектре представляют скорее математический, а не физический интерес, поскольку  при ультрарелятивистских скоростях заряда возрастает роль высокочастотной области спектра,  в которой сделанное в настоящей работе предположение об идеальной проводимости стенок волновода становится
     необоснованным (см., например, [11]).

  %\newpage
\vspace{1cm}
    \begin{center}
    \large\textbf{4. Потери энергии заряда на возбуждение поля}
    \end{center}

     Возбуждение поля в КДВ происходит за сч\"ет потери зарядом энергии. Ввиду бесконечной протяж\"енности структуры и предположения о постоянстве скорости заряда следует считать его массу и энергию бесконечной. В рассматриваемой задаче имеет смысл говорить о потере энергии при прол\"ете зарядом одного периода КДВ. Эту потерю можно найти двумя способами:   либо вычислить изменение энергии электромагнитного поля во всей структуре за время прол\"ета зарядом одного периода КДВ,
     либо найти работу возбуждаемого поля над зарядом  вдоль его траектории заряда в рассматриваемой ячейке.
     Начн\"ем с первого способа.

     Рассмотрим изменение энергии возбужд\"енного в КДВ поля, начиная с момента
     \[t=t_0=-\frac d v+\frac {2D} v N\,,\] за интервал времени \(\Delta t=2D/v\), то есть за время прол\"ета
     зарядом одного периода структуры. Будем считать, что \(N\gg 1\), и, следовательно, поле в
     рассматриваемой ячейке и всех ячейках, расположенных  левее не\"е,  представляет собой
     суперпозицию собственных волн КДВ дискретного спектра, описываемых  формулами (2.5) и (2.7),
     в которых амплитуда \(\mathcal{A}_s\) определена формулой (3.5), \(\psi=2D\omega_s/v\), а \(\theta_s=\pi/2\).
      Разделим условно структуру на две части: справа и слева от плоскости \(z=-d\).  Изменение энергии поля слева от этой плоскости равно потоку через не\"е вектора Пойнтинга \(\Delta P\) за интервал \(\Delta t\). При \(\Delta P<0\)  эту энергию следует  считать <<утекающей>>  на бесконечность, при \(\Delta P>0\) --- <<притекающей>> из бесконечности,  поскольку энергия поля в любой ячейке этой области ввиду постоянства амплитуд возбужд\"енных  собственных волн в среднем по времени не изменяется.
     Изменение энергии поля справа от  плоскости \(z=-\,d\) находится следующим образом. Поскольку  поле,
     которое в момент \(t=t_0\) существовало правее этой плоскости переносится на один период КДВ и становится  расположенным правее плоскости \(z=2D-d\),  то вс\"е изменение энергии поля этой части структуры обусловлено появлением новой порции поля, заполнившего за интервал времени \(\Delta t\) рассматриваемую ячейку.
     Таким образом, полное изменение энергии поля во всей структуре \(\Delta W\)  может быть записано в виде
     \[\Delta W= \overline{W}-\Delta P=\sum_{s=1}^\infty (\overline{W}_s - 2\frac D v\overline{P}_s)\,,\eqno(4.1)\]
     где \(\overline{W}\) -- средняя по времени энергия поля в рассматриваемой ячейке , а запись в виде суммы по спектру возбужд\"енных собственных волн обусловлена тем, что из-за ортогональности их полей  и полный поток вектора Пойнтинга \( \Delta P\) за время \(\Delta t\) через плоскость \(z=-\,d\),
     и   энергия поля в ячейке \(\overline{W}\)  могут быть представлены в виде суммы парциальных вкладов.

        Перейд\"ем теперь к вычислению работы тормозящей заряд силы, обусловленной
     непосредственным действием   возбужд\"енного поля на заряд по мере его движения вдоль КДВ.
     Для этого необходимо  знать составляющую поля   \(E_z\) в точке \(z\) в момент времени \(t=z/v\).
     Работа силы реакции излучения \(T\) при прол\"ете зарядом рассматриваемой ячейки выражается интегралом
     \[T=Q\int\limits_{-d}^{2D-d}E_z(\rho=0,z,t=z/v)\,dz\,.\eqno(4.2)\]
     Основная трудность вычисления этого интеграла  в том, что сама составляющая \(E_z\)
     представляется в виде интегралов Фурье (3.1), (3.2), прич\"ем в данном случае интерес
     представляют те моменты времени, для которых вычисление интеграла пут\"ем перехода в
     комплексную плоскость \(\omega\)  и использование теории  вычетов нецелесообразны,
     поскольку интеграл по дуге большого радиуса не равен нулю и, кроме того,
     нельзя пренебречь вычетами в полюсах на мнимой оси. Поэтому приходится проводить
     интегрирование по действительной оси   \(\omega\).
     Изменив порядок интегрирования по \(\omega\) и \(z\), получаем в результате интегрирования по \(z\):

     \[\left.\begin{array}{l}
     \displaystyle{T=2iQ\int\limits _{-\infty}^\infty\,d\omega\sum_{n=1}^\infty\frac{\nu_n}{J_1(\nu_n)}}
     \times\\[0.6cm]
     \times\Big\{\displaystyle{\frac
     {\Big[F_n\Big(\frac \omega v\sin\frac \omega vl-h_n^a\tg h_n^a l\cos\frac \omega vl\Big)\!-
              G_n\Big(\frac \omega v\cos\frac \omega vl-h_n^a\ctg h_n^al\sin\frac \omega vl\Big)\Big]}
     {(\Gamma a)^2+\nu_n^2}}\,+\\[0.6cm]
      +\displaystyle{\frac {\Big[D_n\Big(\frac \omega v \sin\frac \omega v d-h_n^b\tg h_n^bd\,\cos\frac \omega v d\Big)\!-
     E_n\Big(\frac \omega v \cos \frac \omega v d-h_n^b\ctg h_n^b d\,\sin\frac \omega v d\Big)\Big]}
     {(\Gamma b)^2+\nu_n^2}}\Big\}\,.
     \end{array}\right\} \eqno(4.3)\]

     Интеграл по \(\omega\) является несобственным не только из-за пределов интегрирования, но также из-за наличия  полюсов у коэффициентов \(D_n,\,E_n,\,F_n,\,G_n,\) соответствующих нулям детерминанта матрицы \(\rv M\),  определяемой формулой (1.19). При отсутствии этих полюсов интеграл (4.3) был бы
     равен нулю, так как \(D_n\) и \(F_n\)  неч\"етные, а \(E_n\)  и \(G_n\) ч\"етные функции \(\omega\).
     Если бы полюсы коэффициентов лежали точно на действительной оси, то интеграл (4.3) был бы расходящимся. Поскольку из-за физически неизбежного поглощения в стенках структуры полюсы на комплексной плоскости \(\omega\) лежат чуть ниже действительной оси, то можно воспользоваться известной формулой
    \[\int\limits _{-\infty}^\infty f(\omega)\,d\omega= V.p.\int\limits_{-\infty}^{\infty}f(\omega)\,d\omega
    -i\pi\sum_{s=1}^\infty Res [f(\omega);\omega_s)]\,,\]
    где \(V.p.\) -- интеграл в смысле главного значения, \(Res[f(\omega);\omega_s)]\) -- вычет функции \(f(\omega\)) в полюсе  \(\omega_s\), лежащем ниже действительной оси и бесконечно близко к ней. Обоснование применимости этой формулы при  вычислении аналогичных интегралов, встречающихся в теории излучении Вавилова-Черенкова, имеется в [10].

    Интеграл в смысле главного значения равен нулю, поскольку  подынтегральное выражение неч\"етная функция \(\omega\). Поведение коэффициентов, например, \(D_n,\,E_n\) вблизи полюсов, описываемое формулой (3.3), позволяет представить  их в виде
     \[D_n(\omega)=\frac{\mathcal{A}_s\,\tilde{D}_{s,n}}{4\pi(\omega-\omega_s)}\,,\quad
       E_n(\omega)=\frac{\mathcal{A}_s\,\tilde{E}_{s,n}}{4\pi(\omega-\omega_s)}\,. \eqno(4.4)\]
    В результате вычисления вычетов работу силы реакции излучения можно записать в виде
    \[T=\sum_{s=1}^\infty T_s\,,\]
    где каждый член суммы \(T_s\) обусловлен вычетами в двух полюсах \(\omega=\pm\omega_s\)  и равен

    \[\left.\begin{array}{l}\displaystyle{T_s=\mathcal{A}_sQ
    \sum_{n=1}^\infty\frac{\nu_n}{J_1(\nu_n)}}\times\\[0.6cm]
    \times\!\Big\{
     \displaystyle{\frac
     {\big[\tilde{F}_{s,n}\big(\frac{\omega_s} v\sin\frac {\omega_s} v l\!-\!h_n^a\tg h_n^al\,\cos\frac {\omega_s}  vl\big)
     -\tilde{G}_{sn}\big(\frac{\omega_s} v\cos\frac {\omega_s}  v l\!-\!h_n^a\ctg h_n^al\,\sin\frac {\omega_s} vl\big)\big]} {(\Gamma_s a)^2+\nu_n^2}}+\\[0.6cm]
     +\displaystyle{\frac{\big[\tilde{D}_{s,n}\big( \frac{\omega_s} v\sin\frac {\omega_s} v d\!-\!
     h_n^b \tg h_n^bd \,\cos\frac{\omega_s} v d\big)
     \!-\!\tilde{E}_{s,n}\big( \frac{\omega_s} v\cos \frac {\omega_s} v d\!-\!h_n^b \ctg h_n^b d\,\sin\frac {\omega_s} v d\big)\big]}{(\Gamma_s b)^2+\nu_n^2}}\!\Big\}.\end{array}\!\right\}\eqno(4.5)\]

     На основании закона сохранения энергии должно выполняться равенство
     \[-T_s= \overline{W}_s-\overline{P}_s\Delta t\,,\eqno(4.6)\]
     где \(\Delta t=2D/v\); его невыполнение свидетельствовало бы об ошибке либо  в аналитическом  решении, либо в компьютерной программе вычислений.

     Большой интерес представляет вычисление суммы потоков вектора Пойнтинга возбужд\"енных собственных волн через плоскость \(z=-d\).
     Из изложенных выше соображений следует, что эта сумма не должна быть положительной. В противном случае  получилось бы, что из области отрицательных значений \(z\), которую заряд давно пролетел и где нет других источников поля, в сторону заряда существует постоянный поток энергии, а энергия поля в этой области не изменяется. Поэтому в этой области КДВ  должен либо преобладать вклад волн с отрицательной групповой скоростью, либо суммарный поток вектора Пойнтинга должен обратиться в нуль. Последняя возможность представляется нам наиболее естественной,  хотя  убедиться в том, что именно она имеет место  при движении заряда в КДВ, можно только пут\"ем численных  расч\"етов   для широкого набора геометрических параметров структуры и энергий заряда.

%\newpage

   \begin{center}
   \large\textbf{5. Алгоритм численных расч\"етов}
   \end{center}

   Привед\"енные в предыдущих разделах формулы для описания электромагнитного поля, возбуждаемого зарядом    в КДВ, исчерпывают аналитическую часть решения задачи. Получение количественных результатов, а также выявление обсуждавшихся выше закономерностей возможно лишь пут\"ем численных расч\"етов, которые ввиду их чрезвычайной  громоздкости принципиально требуют использования компьютера. Программа вычислений, результаты которых приведены в следующем разделе \textbf {6},  написана на языке ФОРТРАН Power Station для IBM PC.

   Алгоритм численного расч\"ета энергетических характеристик излучения включает в себя следующую
   последовательность операций: 1) поиск  последовательных нулей \(\omega_s\) определителя матрицы
   \(\rv M\) (1.19) при набеге фазы \(\psi\) (1.14); 2) вычисление для найденных нулей \(\omega_s\)
   элементов матрицы \(\tilde{\rv M}(\omega_s)\) (2.4), 1-й столбец которой представляет собой коэффициенты
   \(\tilde{D}_{s,n}\) и \(\tilde{E}_{s,n}\), а 1-я строка -- вектор \(\bf b\); 3) вычисление амплитуд \(\mathcal{A}_s\) согласно формуле (3.5); 4) расч\"ет работы поля излучения  \(T_s\) (4.5) при прол\"ете зарядом одной ячейки КДВ, потока  мощности \(\overline{P}_s\) (2.9) за время \(\Delta t=2D/v\,,\) энергии поля \(\overline{W}_s\) в рассматриваемой  ячейке КДВ при \(t\to\infty\) и  групповой скорости \(v_{gr}\)  собственной волны, определяемой согласно соотношению (2.13).

   Основная цель настоящей работы состоит в расч\"ете потерь заряда на излучение при прол\"ете им одного периода КДВ в зависимости от его релятивистского фактора \(\gamma\,,\) который определяет верхнюю границу частот возбуждаемых собственных волн. Нижняя граница  частот от \(\gamma\)
   не зависит и совпадает с низшей критической частотой регулярного круглого волновода с радиусом \(b\),
   то есть  \(k_{min}b\geqslant \nu_1\approx 2,405\). Верхняя граница частот \(\omega_{max}\), выше которой амплитуды возбуждаемых волн убывают по экспоненциальному закону, примерно определяется условием \(k_{max}a\sim 3\gamma.\) Расчет потерь на излучение следует начинать с нахождения числа собственных волн \(N_s\) рассматриваемого КДВ, имеющих место в этой полосе частот, поскольку все эти волны возбуждаются и дают вклад в потери.

   Проще всего это число установить при фиксированных значениях набега фазы \(\psi=0\) и\(\psi=\pi\),
   вычислив число нулей определителей матриц \(\bf{P,Q,R,S}\), введ\"енных формулой (1.20). Суммарное число
   нулей для матриц \(\bf{Q}\) и \(\bf{R}\),  соответствующих частотам собственных волн при  \(\psi=0\), должно совпадать с суммарным числом нулей для матриц \(\bf{P}\) и \(\bf{S}\), соответствующих частотам собственных волн при  \(\psi=\pi\). Поиск нулей осуществляется последовательным вычислением определителя матрицы, начиная с \(\omega_{min}\),  с некоторым шагом \(\Delta\omega\). Если при увеличении частоты на один шаг изменяется знак определителя соответствующей матрицы, то в этом интервале находится неч\"етное число собственных волн КДВ. После этого следует уменьшить шаг и
   убедиться, что в уменьшенном интервале частот находится только один нуль определителя. Дальнейшее
   уточнение положения нуля производится по интерполяционной формуле. Если первоначальный шаг
   \(\Delta\omega\) выбран слишком большим, то между двумя последовательными частотами может оказаться два или большее ч\"етное число нулей, и все они будут пропущены, поскольку определитель не сменит знак.
   Слишком маленькое значение шага неоправданно увеличивает время сч\"ета. Для каждого конкретного КДВ существует свой оптимальный шаг. Чтобы не пропустить ч\"етное число собственных волн, целесообразно провести вычисление числа смены знаков определителя с уменьшенным в несколько раз шагом. И только убедившись, что число этих смен не изменилось, уточнять положение нулей.

  Следующим этапом вычислений для каждого нового варианта КДВ является построение дисперсионных кривых собственных волн в диапазоне набега фазы \(\psi\) от 0 до \(\pi\). При этом необходимо убедиться, что дисперсионные кривые не пересекаются, и каждая кривая, исходящая из точки \(\omega_s\) на оси \(\psi=0\), попадает в точку \(\omega_s\) на линии \(\psi=\pi\) с тем же последовательным номером \(s\).
   Для всех значений \(\psi,\) для которых строится дисперсионная кривая, необходимо проверить выполнение
   равенств    \(\overline{P}_s^A= \overline{P}_s^B\) и \(\overline{W}_s^H=\overline{W}_s^E\), а также совпадение значений групповой скорости \(v_{gr}\), вычисленных в соответствии с формулами (2.1) и (2.13). Кроме того, необходимо убедиться, что выполняется неравенство \(v_{gr}<c\).

   Для вычисления энергетических характеристик собственных волн КДВ, таких как
   \(\overline{P}_s^A, \overline{W}_s^H\), необходимо найти элементы матрицы \(\tilde{\rv M},\)
    определ\"енной формулой (2.4). Для этого с помощью стандартной процедуры обращения матрицы  вычисляем матрицу \(\rv M^{-1}\) на частоте \(\omega=\omega_s(1+\varepsilon)\), на которой матрица
    \(\rv M\) уже невырожденная. Вычисления показывают, что при изменении параметра \(\varepsilon\)
     в пределах от \(10^{-6}\)  до \(10^{-12}\) у всех элементов матрицы \(\tilde{\rv M}\) первые четыре цифры остаются неизменными. Оптимальным значением этого параметра следует считать \(10^{-9}\); оно определяется тем, что в программе вычислений используется тип переменных DOUBLE PRECISION.

     После того, как для данного КДВ рассчитана структура дисперсионных кривых собственных волн в требуемом диапазоне частот, можно переходить к расчету потерь на излучение для заданного значения
     энергии заряда. Процедура вычисления частот \(\omega_s\) и матриц \(\tilde{\rv M}(\omega_s)\)такая же, как описано выше, за исключением того, что в этом случае переменные \(\psi\) и \(\omega\) связаны соотношением (1.14). Амплитуды \(\mathcal{A}_s\)  и потери на излучение вычисляются по приведенным выше аналитическим формулам.

     Все численные расч\"еты производятся для безразмерных переменных. Вместо частоты \(\omega\)
     используется переменная \(kb\), величины \(T_s,\) \(\overline{W}_s\) и \(\overline{P}_s\Delta t\) нормируются на множитель \(Q^2/b\). Для всех численных результатов следующего раздела подразумевается наличие этого нормирующего множителя.

     %\newpage
    \vspace{1cm} \begin{center}
     \large\textbf{6. Основные результаты вычислений}
     \end{center}

  Все приводимые ниже результаты расчетов дисперсионных кривых собственных волн и потерь энергии точечного заряда на излучение   относятся к КДВ, геометрические размеры которого совпадают с
  размерами периода структуры ускоряющих секций    Станфордского линейного ускорителя SLAC [10]:
  \[b = 4,173 \;\textrm{см},\quad a = 1,321\;\textrm{см},\quad D = 1,750 \;\textrm{см},\quad d =1,458 \;\textrm{см}.\]

На рис. 2 представлены результаты расч\"ета числа собственных волн КДВ \(N_s\) во вс\"ем диапазоне частот вплоть до частоты, определяемой значением безразмерной переменной \(kb\).

В нижеследующей таблице  1 приводятся результаты вычислений безразмерных величин \(k_sb\) и соответствующих частот \(f_s\) (в МГц) собственных волн  для набегов фаз \(\psi=0\) и \(\psi=\pi\).  Фактически вычислялись собственные частоты вспомогательных резонаторов, представляющих собой половину ячейки КДВ с соответствующими торцевыми стенками (см. раздел \textbf{2}).
 
Дисперсионные кривые собственных волн  КДВ на плоскости переменных \(kb,\psi/\pi\) строятся, начиная с
уже вычисленного значения \(k_sb(\psi=0)\),  с некоторым шагом \(\Delta \psi\). Приближ\"енное значение
\(k_s b\) на следующем шаге находится с учетом соотношения (2.1) по формуле
\[k_s b(\psi+\Delta \psi)= k_s b(\psi)+\beta_{gr}\frac {b\Delta \psi} {2D}\,.\]
Уточнение этого значения производится методом поиска нуля определителя матрицы \(\rv M\) (1.19) пут\"ем его вычисления с некоторым шагом \(\Delta (kb)\) по обе стороны от приближ\"енного значения. Для волн с большим номером \(s\) оба расч\"етных параметра  \(\Delta \psi\) и  \(\Delta (kb)\) приходится выбирать достаточно малыми, иначе возможно получение в качестве уточн\"енного значения точки соседней дисперсионной кривой. Только при правильном выборе этих параметров дисперсионная кривая попад\"ет в предварительно вычисленную точку \(k_sb(\psi=\pi)\) с тем же номером \(s\).

 \begin{center}
 \includegraphics[width=0.8\textwidth,draft=false]{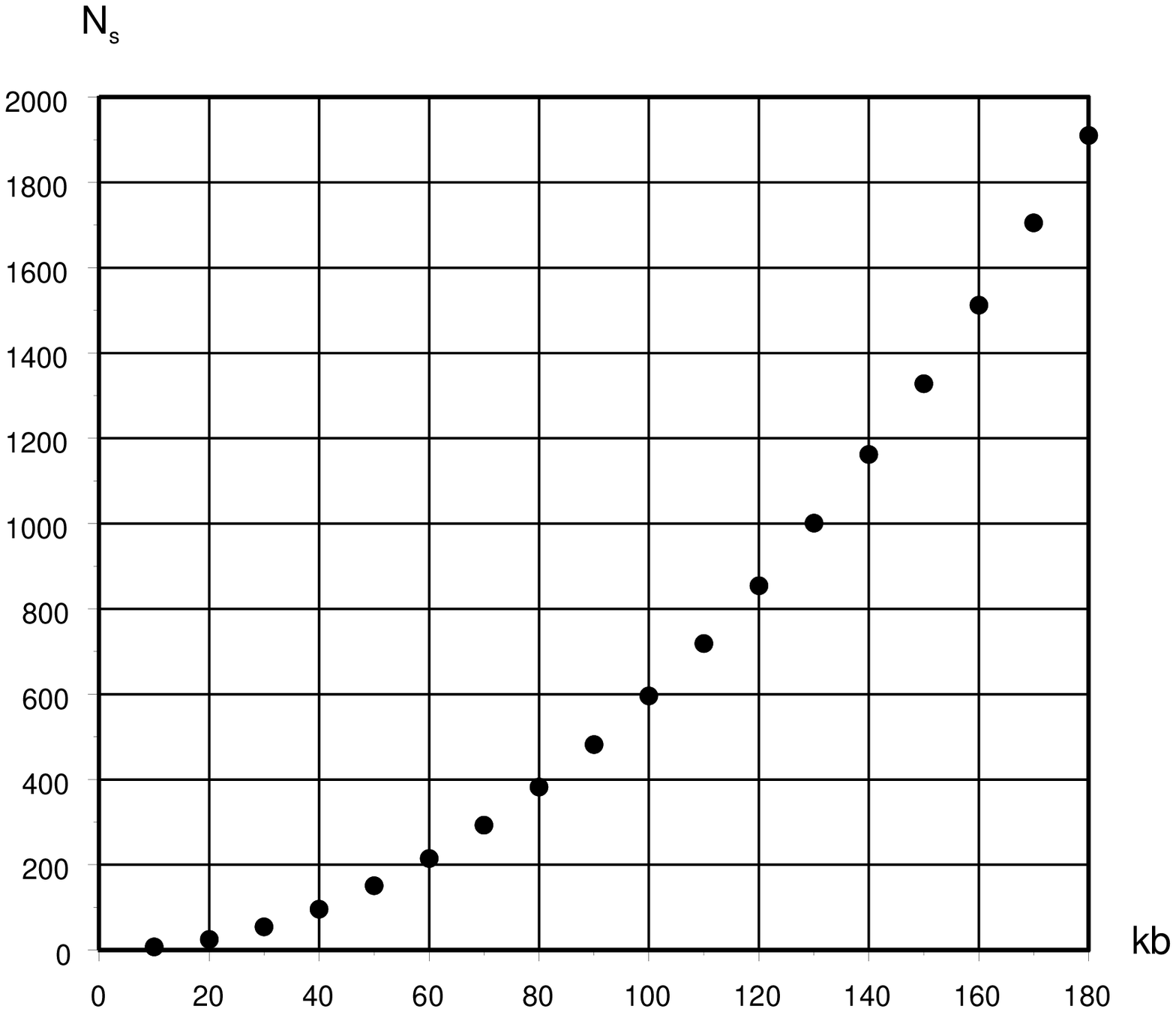}
  \end{center}
 \hspace{25 mm} Рис. 2. Число собственных волн КДВ в функции \(kb\)
\vspace{5 mm}
 %\newpage

\begin{center}\textbf
{Таблица 1\\ Результаты вычислений \(k_sb\) и соответствующих частот \(f_s\)}
\end{center}
  \begin{center}
  \begin{tabular}{|r||c|c||c|c|}\hline
  &\multicolumn{2}{|c||}{\(\psi = 0\)}&\multicolumn{2}{|c|}{\(\psi = \pi\)} \\ \hline
  s & \(k_s b\) & \(f_s, \textrm{МГц} \)& \(k_s b\) &\( f_s,\textrm{МГц} \)\\ \hline
  1 & 2,4456 & 2796,2& 2,4917 & 2849,0 \\ \hline
  2 & 5,1711 & 5912,5 & 5,1261 & 5861,1 \\ \hline
  3 & 5,6032 & 6406,6 & 5,8159 & 6649,9 \\ \hline
  4 & 7,5570 & 8640,6 & 7,1628 & 8189,9 \\ \hline
  5 & 8,5820 & 9812,6 & 8,9266 & 10206,5  \\ \hline
  \ldots&&&&\\ \hline
  86 & 37,4409 &42809,3 & 37,4096 & 42773,6 \\ \hline
  87 & 37,8457 & 43272,2 & 38,0529 & 43509,1 \\ \hline
  88 & 37,8887 & 43321,3 & 38,1782 & 43652,4 \\ \hline
  89 & 38,2611 & 43747,1 & 38,3477 & 43846,2 \\ \hline
  90 & 38,4217 & 43930,8 & 38,5067 & 44028,0 \\ \hline
  \ldots &&&&\\ \hline
  \end{tabular}
  \end{center}

Для каждого вычисленного значения  \(k_s b(\psi)\) следует убедиться, что выполнены все перечисленные выше равенства для энергетических величин собственных волн. При порядке редукции \(N=250\) матрицы \(\rv M\) (1.19) относительная ошибка в равенстве \(\overline{P}_s^A=\overline{P}_s^B\) для всех волн с номерами \(s<2000\) не превышает \(10^{-5}\) (за исключением окрестностей точек, в которых \(v_{gr}=0\)). Для равенства
\(\overline{W}_s^H=\overline{W}_s^E\) относительная ошибка \(<10^{-6},\) несовпадение значений \(v_{gr}\), вычисленных двумя разными способами не превышает \(10^{-4}\).

Для первых 18  волн имеет место хорошо известный вид  дисперсионных кривых.  Каждая кривая в интервале \(\psi\)  от 0 до \(\pi\) представляет собой монотонную функцию, что обеспечивает постоянство знака групповой скорости. Для нескольких первых волн  ширина полосы частот, в пределах которой  существуют распространяющиеся вдоль КДВ волны, существенно меньше, чем расстояние до следующей полосы прозрачности. Однако уже между 16-й и 17-й волнами полоса частот, где волны не могут распространяться, отсутствует, хотя  дисперсионные кривые для обеих волн представляют собой монотонные функции.  Далее с ростом номера волны картина существенно усложняется.  Для ряда кривых появляются одно или более значений \(\psi\), при которых  имеет место локальный максимум или минимум, групповая скорость в этой точке обращается в нуль, и на соответствующей частоте может быть возбуждена стоячая волна. При переходе через эту точку групповая скорость собственной волны меняет знак. Появляются широкие интервалы частот,
в пределах которых одна из волн, а то и одновременно несколько соседних, являются распространяющимися,   причем на одной и той же частоте могут быть возбуждены волны с разными  номерами и разными знаками групповой скорости. Пример такого поведения нескольких последовательных дисперсионных кривых представлен на рис. 3.

Как видно из рисунка, между 86-й и 87-й волнами оста\"ется узкий интервал частот, в котором распространяющихся волн нет.  С дальнейшим ростом номера собственной волны распределение дисперсионных кривых становится ещ\"е более причудливым, интервалы "непрозрачности" \,  встречаются   реже, и они становятся вс\"е более узкими.

  При вычислении потерь энергии заряда, движущегося вдоль КДВ, в зависимости от его релятивистского фактора \(\gamma\), определяющего значение \(k_{max}b\), знание соответствующего числа собственных  волн   \(N_s\) и вида их дисперсионных кривых позволяет правильно выбрать исходное
  значение важного  расч\"етного параметра \(\Delta kb\) -- шага, с которым производится поиск числа интервалов,  в пределах которых определитель матрицы \(\rv M\) меняет знак. Если число таких интервалов оказывается меньшим \(N_s\), то величину шага \(\Delta kb\) необходимо уменьшить. Уже при \(\gamma = 20\) верхний предел \(kb\) составляет 190, чему соответствует \(N_s=2220\). При этом шаг  \(\Delta kb\) не должен превышать 0,002. Хотя при порядке редукции матрицы \(\rv M\;N=500\), который использовался при получении всех привед\"енных  далее результатов, точность вычисления энергетических характеристик излучения оста\"ется приемлемой для всех собственных волн с номерами \(s<2500\), требуемое число операций ограничивает возможность дальнейшего увеличения \(\gamma = 20\) при параметрах компьютера, на котором проводились расч\"еты.

   Если построены дисперсионные кривые собственных волн в требуемом интервале частот, то графически нетрудно определить частоту, на которой возбуждается  волна заданного номера: это точка пересечения соответствующей кривой рис. 3 с прямой линией \(k_s b=\psi\beta b/2D\) (при \(\psi=1\) прямую следует зеркально отразить,  далее последовательно отразить при \(\psi=0\) и так далее).  На рис. 3 изображены две такие линии: линия для \(\beta=1\), определяющая частоты предельного спектра, и линия для \(\gamma=3,5\).

  \begin{center}
  \includegraphics[width=0.95\textwidth,draft=false]{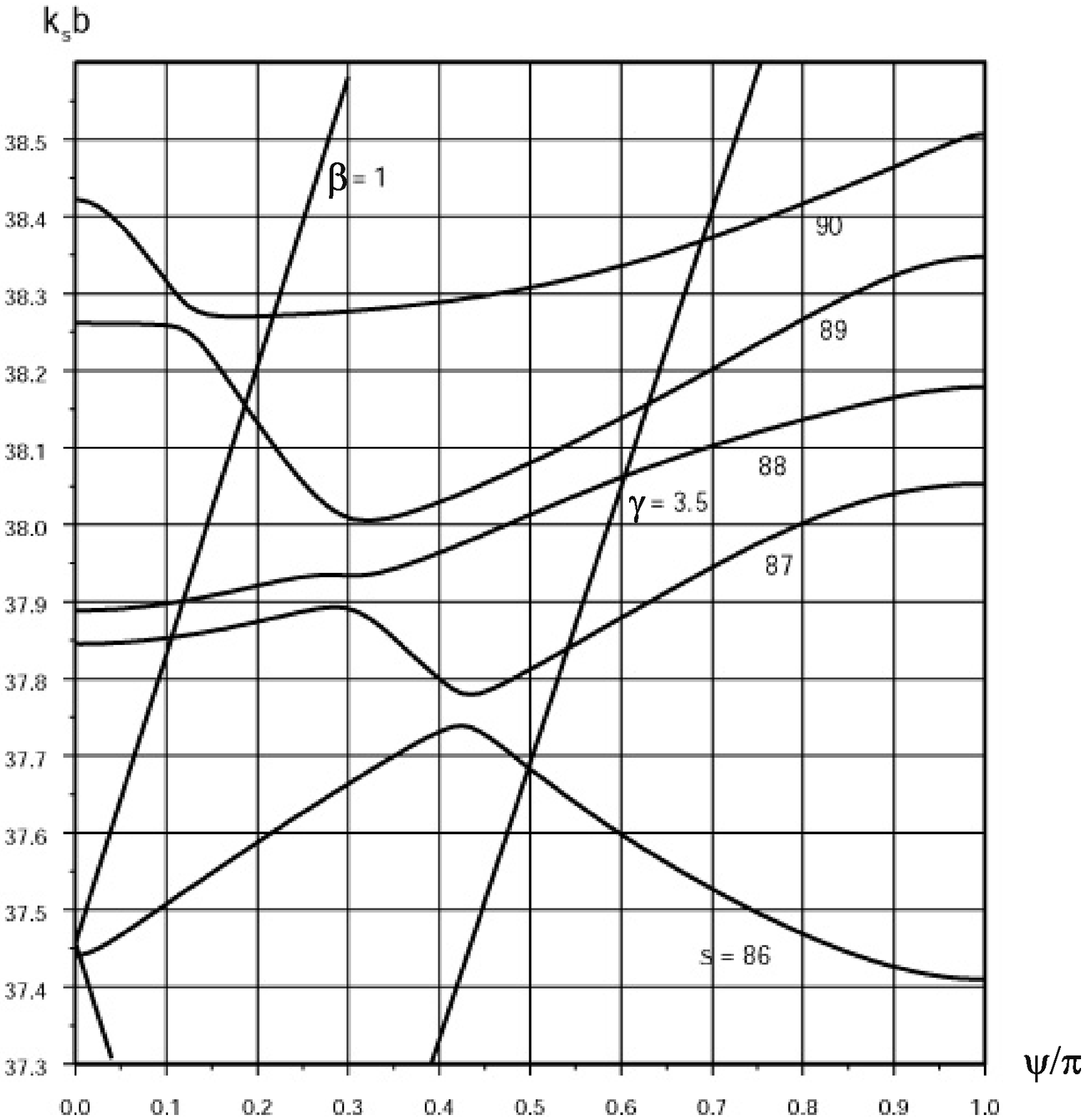}
    \end{center}
  \hspace{25mm}   Рис. 3. Дисперсионные кривые собственных волн КДВ
  \vspace{5 mm}

   В таблице 2 приводятся результаты расч\"ета энергетических характеристик поля излучения заряда при \(\gamma=3,5\). Следует отметить, что равенство (4.6) выполняется для всех собственных волн с точностью порядка десятых долей процента.
   Этот  факт  является определ\"енной гарантией правильности всего подхода к задаче и отсутствия ошибок как в аналитических выкладках,
   так и программе вычислений, поскольку одна из величин, входящих в это равенство, пропорциональна амплитуде собственной волны \(\mathcal{A}_s\),
   а две другие -- квадрату амплитуды. Отметим также, что работа поля излучения над зарядом для всех волн имеет один и тот же правильный знак.
   \vspace*{-7.5mm}
  \begin{center}\textbf
 {Таблица 2\\ Энергетические характеристики поля излучения заряда} (\(\gamma=3,5\))
 \end{center}
 \begin{center}
 \begin{tabular}{|r|c|c|c|c|c|c|c|c|c|}\hline
 s&\(k_sb\)&\(\psi/\pi\)&\(\beta_{gr}\)&\(\overline{W}_s\)&\(T_s\)&\(\overline{P}_s\Delta t\)
 &\(\sum \overline{W}_s\) &\(\sum T_s\)&\(\Delta t\sum \overline{P}_s\) \\ \hline
 %&&&&\(Q^2/b\)&\(Q^2/b\)&\(Q^2/b\)&\(Q^2/b\)&\(Q^2/b\)&\(Q^2/b\)\\ \hline
 1&2,482&0,691&0,016&2,8692&-2,8196&0,0476&2,8692 &-2,8196&0,0476\\ \hline
 2&5,141&1,432&0,017&0,4286&-0,4201&0,0077&3,2978&-3,2398&0,0552\\ \hline
 3&5,678&1,582&-0,083&0,2774&-0,3016&-0,0240&3,5752&-3,5413&0,0312\\\hline
 4&7,546&2,102&-0,056&0,6762&-0,7148&-0,0397&4,2513&-4,2561&-0,0085\\ \hline
 5&8,712&2,427&0,134&0,0497&-0,0428&0,0069&4,3010&-4,2989&-0,0015\\ \hline
 \(\ldots\)&&&&&&&&&\\ \hline
 45&27,300&7,605&-0,183&0,0060&-0,0071&-0,0011&5,5088&-5,5075&-0,0031\\ \hline
 46&27,501&7,661&-0,027&0,0000&0,0000&0,0000&5,5088&-5,5075&-0,0031\\ \hline
 47&27,747&7,730&-0,012&0,0075&-0,0075&-0,0001&5,5163&-5,5150&-0,0032\\ \hline
 48&27,855&7,760&0,004&0,0005&-0,0005&0,0000&5,5170&-5,5155&-0,0032\\ \hline
 49&28,248&7,869&0,114&0,0004&-0,0004&0,0000&5,5171&-5,5159&-0,0032\\ \hline
 \(\ldots\)&&&&&&&&&\\ \hline
 86&37,684&10,498&-0,245&0,0004&-0,0005&-0,0001&5,5487&-5,5474&-0,0034\\ \hline
 87&37,840&10,542&0,179&0,0014&-0,0012&0,0003&5,5501&-5,5486&-0,0032\\ \hline
 88&38,063&10,604&0,120&0,0000&0,0000&0,0000&5,5501&-5,5486&-0,0032\\ \hline
 89&38,157&10,630&0,167&0,0001&-0,0001&0,0000&5,5502&-5,5487&-0,0031\\ \hline
 90&38,368&10,689&0,108&0,0002&-0,0002&0,0000&5,5504&-5,5489&-0,0031\\ \hline
 \end{tabular}

 %\end{center}
 % \begin{center}
  %\includegraphics[bb= 0mm 0mm 180mm 115mm,scale=0.85,draft=false]{T}
 \includegraphics[width=0.98\textwidth,draft=false]{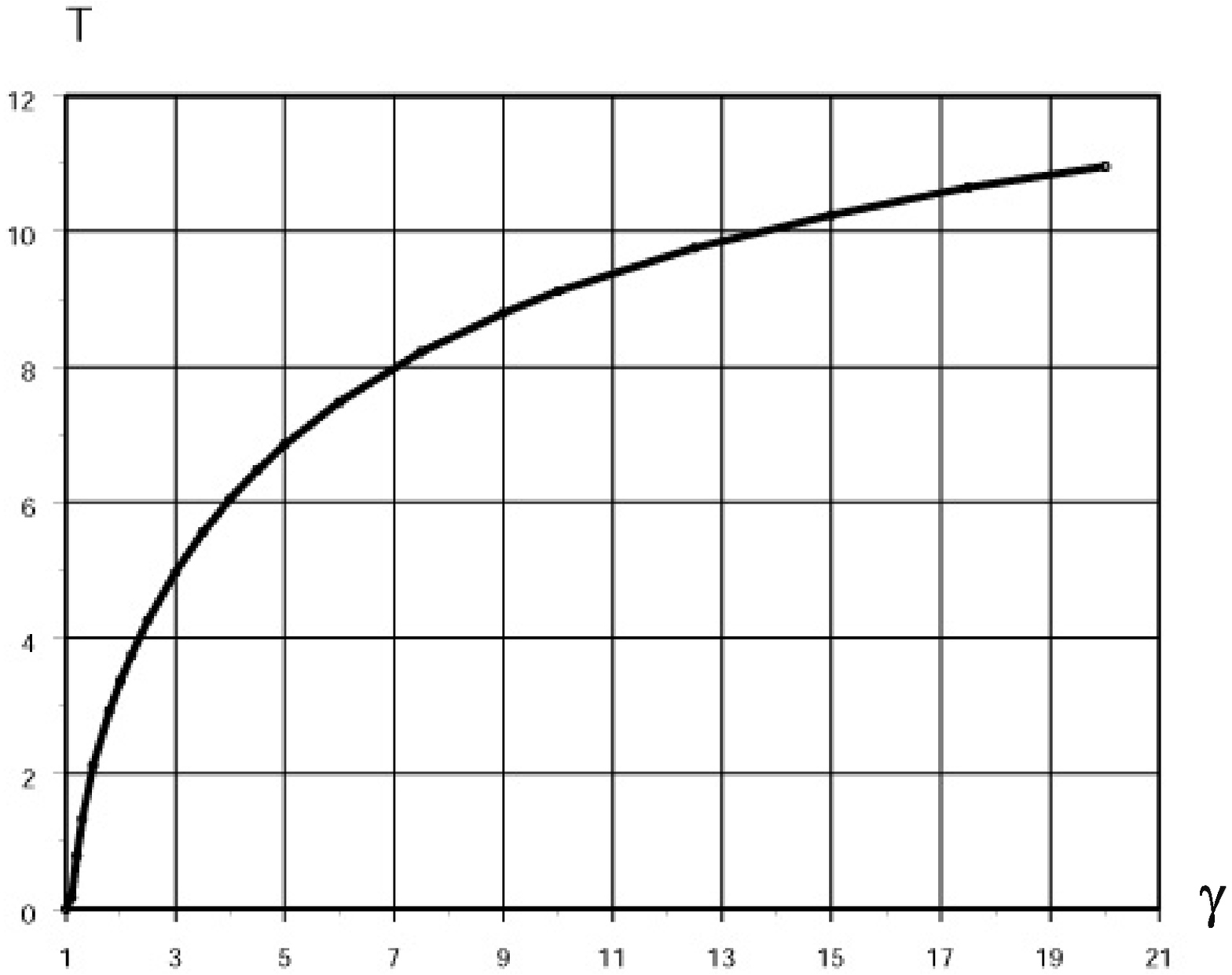}
   \end{center}
  \vspace{-5mm}\hspace{17 mm} Рис. 4. Потери энергии зарядом на излучение на периоде КДВ
 \vspace{1cm}

Направление вектора Пойнтинга, определяемое знаком групповой скорости собственной волны, у одних волн совпадает с направлением движения заряда, а у других -- противоположно ему. Суммарный поток всех волн мал по сравнению с суммарной работой поля излучения и, возможно, находится в пределах
   погрешностей вычислений.

 Значения \(k_s b\) для волн с номерами \(s=86-90\)  совпадают с точками пересечения соответствующих дисперсионных кривых
 и прямой линии для \(\gamma=3,5\) на рис. 3.
  %\newpage

 На рис. 4 представлена зависимость потерь энергии зарядом на излучение при прол\"ете им периода КДВ от его релятивистского фактора \(\gamma\).

 Для наибольшего значения \(\gamma=20\), представленного на графике, был учт\"ен вклад 2500 собственных волн. Уч\"ет этого же числа волн для
 \(\gamma=10^8\) да\"ет для работы поля излучения над зарядом на периоде структуры значение \(T=12,90\).  Очевидно, что для всех учт\"енных волн вклад в потери был такой же, как и в случае предельного спектра, соответствующего \(\gamma\to \infty\). Поскольку вклад последних 200 учт\"енных волн составил менее 0,005 вклада всех предыдущих, то можно утверждать, что если расходимость потерь предельного спектра и имеет место, то она очень медленная. Более вероятным представляется стремление потерь к постоянному значению с ростом \(\gamma\).

  %\newpage

  \vspace{1cm}\begin{center}
  \large\textbf{Заключение}
  \end{center}

  Аналитическое решение задачи, основанное на трансляционной симметрии поля, возбуждаемого  зарядом при прол\"ете   каждого периода КДВ, определяет простран-  ственно-временную   структуру возбужд\"енного
  поля: впереди заряда  поле экспоненциально убывает с расстоянием от заряда и незначительно отличается от поля заряда, движущегося в  регулярном круглом волноводе; в области структуры, которую заряд давно пролетел, поле представляет собой совокупность  собственных волн КДВ.  Амплитуда, частота и набег фазы на периоде собственной волны определяются скоростью заряда. При малых скоростях амплитуды экспоненциально убывают при уменьшении скорости. Уже при \(\beta=0,4\) излучением можно пренебречь. При \(\gamma \to \infty\) все параметры каждой собственной волны стремятся к конечным значениям, определяющим предельный спектр излучения.

  Результаты численных расч\"етов показывают, что работа поля излучения над зарядом при прол\"ете им периода КДВ равна  изменению энергии поля в структуре за это время. Энергетический баланс выполняется не только для суммарного поля, но и для каждой собственной волны. Изменение средней по времени энергии поля волны за сч\"ет увеличения  занятого им объ\"ема структуры связано со средним  потоком вектора Пойнтинга и групповой скоростью волны соотношением, известным из общей теории периодических структур. Среди возбужд\"енных  собственных волн, дающих заметный вклад в суммарное поле, при всех скоростях заряда имеются волны как с положительной,  так и с отрицательной групповой скоростью. В результате суммарный поток энергии вдоль КДВ всех волн за время прол\"ета одного периода оказывается существенно меньшим суммарной работы поля излучения над зарядом. Достигнутая  точность расч\"етов позволяет сделать предположение о строгом равенстве нулю суммарного потока энергии вдоль КДВ.  При этом авторы полностью отдают себе отч\"ет в том, что получить строгое доказательство обращения в нуль какой-бы то ни  было физической величины пут\"ем численных расч\"етов невозможно.

  С ростом энергии заряда происходит насыщение амплитуд нижней части спектра собственных волн на уровне предельного  спектра, соответствующего \(\gamma \to \infty\). Граница верхней части спектра, начиная с которой имеет место экспоненциальное спадание амплитуд собственных волн, сдвигается в сторону высоких частот пропорционально \(\gamma\). Амплитуды собственных волн предельного спектра спадают в среднем с номером волны, что не позволяет сделать вывода о расходимости энергии излучения в предельном спектре. Приведенная на графике кривая потерь энергии заряда на излучение при прол\"ете одного периода КДВ в функции релятивистского фактора заряда показывает медленное нарастание потерь с ростом \(\gamma\), что свидетельствует либо о сходимости энергии излучения предельного спектра, либо об очень слабой расходимости.
%\newpage
   \vspace{1cm}\begin{center}
  \large\textbf{ Список литературы}
  \end{center}

   1.  Е.И. Буляк, В.И. Курилко, В.Г. Папкович. Методы аналитической теории дисперсии круглого
   диафрагмированного волновода (обзор). М., ЦНИИАтоминформ, 1989.
  %\vspace{0.5cm}

   2.   А.Г. Трагов.  Ускорители. Вып. 3. М., Госатомиздат,  1962, с. 148.
  %\vspace{0.5cm}

   3.   П.Е. Краснушкин, С.П. Ломнев.  Радиотехника и электроника, 1966, т.11, №6, с. 1051.
  %\vspace{0.5cm}

   4.  Э.Л. Бурштейн, Г.В. Воскресенский.  Линейные ускорители электронов с интенсивными пучками.
        М.,  Атомиздат,  1970.
 %\vspace{0.5cm}

   5.  E. Keil.  Труды ХII Международной конференции по ускорителям заряженных частиц высокой энергии,
   Ереван, 1970, т.II, c.551.
  %\vspace{0.5cm}

   6.  Л.А. Вайнштейн.  Электромагнитные волны. М., Радио и связь, 1988.
   %\vspace{0.5cm}

   7.  Л.А. Вайнштейн, В.Ф. Солнцев.  Лекции по сверхвысокочастотной электронике. М., Сов. радио, 1973.
    %\vspace{0.5cm}

   8.  В.А. Бережной, В.Н. Курдюмов. Лекции по высокочастотной электродинамике. М., ИЯИ РАН, 2001.
    %\vspace{0.5cm}

   9.  А.Г. Курош.  Курс высшей алгебры. М., ГИТТЛ, 1975.
    %\vspace{0.5cm}

  10.  Б.М. Болотовский. УФН, 1957, т.62, с.201
  %\vspace{0.5cm}

  11. В.А. Бережной, В.Н. Курдюмов. Кильватерные потенциалы и импедансы связи в однородной камере
  с конечной проводимостью стенок. Препринт ИЯИ-1095/2003, М., ИЯИ РАН, 2003.
  %\vspace{0.5cm}

 12.  R.B. Neil.  The Stanford two-mile accelerator. New York - Amsterdam, W.A. Benjamin, Inc., 1968.
     %\vspace{0.5cm}

   \end{document}